\documentclass[english]{article}
\usepackage[T1]{fontenc}
\usepackage[latin9]{inputenc}
\usepackage{textcomp}
\usepackage{amsmath}
\usepackage{graphicx}

\makeatletter

\providecommand{\tabularnewline}{\\}

\newcommand{\lyxaddress}[1]{
\par {\raggedright #1
\vspace{1.4em}
\noindent\par}
}

\usepackage{subfigure}

\makeatother

\usepackage{babel}
\begin{document}

\title{Image Reconstruction Image reconstruction by using local inverse
for full field of view.}

\author{Kang Yang$^{1}$, Kevin Yang$^{1}$, Xintie Yang$^{2}$, Shuang-Ren
Zhao$^{1}$ }

\maketitle

\lyxaddress{$^{1}$Imrecons.com North York, Ontario, Canada}

\lyxaddress{$^{2}$Northwestern Poly Technical University Xi'an China}
\begin{abstract}
The iterative refinement method (IRM) has been very successfully applied
in many different fields for examples the modern quantum chemical
calculation and CT image reconstruction. It is proved that the refinement
method can create a exact inverse from an approximate inverse with
a few iterations. The IRM has been used in CT image reconstruction
to lower the radiation dose. The IRM utilize the errors between the
original measured data and the recalculated data to correct the reconstructed
images. However if it is not smooth inside the object, there often
is an over-correction along the boundary of the organs in the reconstructed
images. The over-correction increase the noises especially on the
edges inside the image. One solution to reduce the above mentioned
noises is using some kind of filters. Filtering the noise before/after/between
the image reconstruction processing. However filtering the noises
also means reduce the resolution of the reconstructed images. The
filtered image is often applied to the image automation for examples
image segmentation or image registration but diagnosis. For diagnosis,
doctor would prefer the original images without filtering process. 

In the time these authors of this manuscript did the work of interior
image reconstruction with local inverse method, they noticed that
the local inverse method does not only reduced the truncation artifacts
but also reduced the artifacts and noise introduced from filtered
back-projection method without truncation. This discovery lead them
to develop the sub-regional iterative refinement (SIRM) image reconstruction
method. The SIRM did good job to reduce the artifacts and noises in
the reconstructed images.

The SIRM divide the image to many small sub-regions. To each small
sub-region the principle of local inverse method is applied. After
the image reconstruction, the reconstructed image has grids on the
border of sub-region inside the object. If do not consider the grids,
the noise and artifacts are reduced compare the original reconstructed
image. To eliminate the grids, these authors have to add the margin
to the sub-region. they did not think the margin is important issue
in the beginning. However when they considering the size of sub-region
tending to only one pixel, they found that the margin play a important
role. This limit situation of sub-regional iterative refinement is
referred as local-region regional iterative refinement(LIRM). If the
margin is very large for example as half the image size, the SIRM
and the LIRM become no iterative refinement method (NIRM), i.e. normal
filtered back-projection method. If the size of the margin tend to
0, the LIRM become the IRM, which is also referred as traditional
iterative refinement method (TIRM). For SIRM and LIRM, if margin is
too small, the image is rich in noise like IRM, if the margin is too
big the image is rich in artifacts like NIRM. If a suitable margin
is taken, for example the margin is around 20 pixel for a 512{*}512
image, the summation of the noise and artifacts can be minimized. 

SIRM and LIRM can be seen as local inverse applied to the image reconstruction
of full field of veiw. SIRM and LIRM can be seen as generalized iterative
refinement method(GIRM). SIRM and LIRM do not minimize the noise or
artifacts but minimize the summation of the noise and artifacts. Even
the SIRM and LIRM are developed in the field of CT image reconstruction,
these authors believe they are a general methods and can be applied
widely in physics and applied mathematics.
\end{abstract}

\section{Introduction}

\subsection{Iterative refinement method (IRM) in applied mathematics, physics
and image recovery / image reconstruction }

The IRM\cite{iterativeRefinement} is widely applied to physics and
applied mathematics for example ref.\cite{W-Moench-Freiberg,Xinyuan-Wu-Jun-Wu}.
The advantage of the IRM is that it can produce an exact inverse from
approximate inverse\cite{Shin-ichi,Robert-Bridsony}. A recent very
important application of the IRM in quantum physics can be found\cite{Anders-M-N-Niklasson}
in which the Eigen values can be calculated dependent linearly with
the size of matrix. The IRM is also applied to image recovery\cite{Haricharan-Lakshman}
and MR image reconstruction\cite{Lin-He}.

\subsection{IRM in CT image reconstruction}

Many reconstruction algorithms have been developed for parallel-beam\cite{Kak-A-C},
fan-beam\cite{NooF,Parker,Kak-A-C,Zhao-S-R-1993,ref_Shuangren_zhaoFourierFan1,ref_Shuangren_zhaoFourierFan2},
and cone-beam tomographic systems\cite{Ref-11-L-A-Feldkamp,Ref-13-B-D-Smith,Ref-14-P-Grangeat,Ref-15-Defrise,Ref-16-Shuangren-Zhao}.
The above reconstruction methods which has no iteration are referred
as direct reconstruction. On the other hand, image can be reconstructed
by IRMs\cite{Chang,Zeng,Riddell,OSullivan,Delaney}. In the IRM, the
reconstructed image is re-projected and the errors between the measured
projections and re-projections are calculated. These errors are utilized
to correct the first reconstruction. In order to distinguish other
IRM which will be introduced later in this article, this IRM is referred
as traditional iterative refinement reconstruction method (TIRM).
The direct reconstruction without iterative refinement is referred
as non iterative refinement reconstruction method (NIRM). The TIRM
is known that it can reduce artifacts especially beam-harden artifacts
at a price of increasing the noises. Moreover, TIRM is more time-consuming
compared to the direct reconstruction (NIRM), since it has a iteration.
Time-consuming in image reconstruction is not a big issue in our generation
since the improvement of the computer hardware. A recent iterative
refinement reconstruction was the work of Johan Sunnegardh\cite{JohanSunnegardh}.
Siemens AG has claimed that Johan's method is their new generation
CT image reconstruction method. This means that big companies begin
to notice the importance of the IRM in image reconstruction. It is
noticed that Johan's iterative refinement reconstruction has a pre-filtering
and post-filtering process which are used to reduce the noises of
their IRM. For CT image reconstruction the IRM can reduce the artifacts
of reconstructed image hence reduce the dose required for CT image.
However the IRM often has an over-corrections on the image edges which
increases the noises. Johan's iterative refinement reconstruction
has to include filtering process to reduce the noises. Pre-filtering
and post-filtering process does not only reduce the noises, but it
also decreases the information contained in the original projection
data. these authors have also roughly introduced a IRM in the past
which is referred as sub-regional iterative refinement method (SIRM)\cite{Ref-25-shuangRenZhao}.
SIRM dose not include pre-fltering and post filtering processes, since
their noise level is not increased. This article offers the details
and also the history how this method has been discovered. The limit
situation when the size of the sub-region approaches to $1$ is also
discussed, in this case the SIRM method becomes local-region iterative
refinement method (LIRM).

\subsection{The history about developing the SIRM and LIRM}

One of the important task in image reconstruction is to reduce the
truncation artifacts. The truncation artifacts are caused by LFOV
(limited field of view) of the detector. The truncation artifacts
can be reduced through extrapolation of the missing rays\cite{Ref-3-Maria-Magnusson-Seger,Ref-8-F-Rashid-Farrokhi,MattiasNilsson}
or local tomography\cite{Ref-7Adel-Faridani,Ref-8-F-Rashid-Farrokhi,Ref-17-Alexander-Katsevich,Ref-18-Alexander-Katsevich,MattiasNilsson}.
Extrapolation often produces an over-correction or an under-correction
to the reconstructed image\cite{Ref-24-Shuangren-Zhao}. Local tomography
cups the reconstructed image. Iterative reconstruction and re-projection
algorithm \cite{Ref-4-Nassi,Ref-5-Paul-S-Cho,Ref-20-J-H-Kim} was
developed for the reconstruction of limited angle of view. Another
iterative reconstruction re-projection algorithm has been developed
for LFOV\cite{Ref-22-Shuangren-Zhao}. The iterative algorithm for
LFOV reduces the truncation artifacts remarkably. This algorithm led
to a truncation free solution local inverse for LFOV\cite{Ref-24-Shuangren-Zhao}.
The fast extrapolation used in \cite{Ref-24-Shuangren-Zhao} is done
in \cite{Ref-23-ShuangrenZhao}. During the work with the iterative
algorithm for LFOV, it was occasionally found that this algorithm
did not only reduce the truncation artifacts but also it reduced the
normal artifacts. Here the normal artifacts means that the artifacts
exist in the direct reconstruction$\:$(NIRM) for example FBP (filtered
back projection) method with full field of view (FFOV). It is well
know that even FBP method is a exact method, it is actually not exact
because the numerical calculations. This finding led to the SIRM for
FFOV\cite{Ref-25-shuangRenZhao}. 

SIRM was adjusted to suit the case of FFOV. This was done by taking
away the extrapolation process. The first reconstruction in the iteration
was FBP method. For the second reconstruction, the region of the object
was divided to many sub-regions. The iterative reconstruction was
done for each sub-region adding a margin. The part of the object from
the 1st reconstruction outside a sub-region (including its margin)
was reprojected. These re-projections were subtracted from the original
projections. The resulted projections were utilized to make the second
reconstruction inside the sub-region. Finally the reconstructed images
on all sub-regions were put together. It is worthwhile to say that
the margin was added in the beginning to eliminate the cracks on the
border of the sub-regions. It was not thought as an important issue.
However it was found that it was the margin that really plays the
role to reduce artifacts \cite{Ref-25-shuangRenZhao} and decreasing
the noises.

\subsection{Other image reconstruction methods}

It is worth to mention that Many fanbeam and conebeam image reconstruction
algorithms have been derived in recent 20 years. Amongst these derivations
are the following examples: the references\cite{key-23,key-24,key-25,key-26,key-40,key-43,key-45,key-49,key-67,key-68,Ref-22-Shuangren-Zhao,Ref-23-ShuangrenZhao,Ref-24-Shuangren-Zhao,Ref-25-shuangRenZhao}
are reconstruction algorithms for truncated projections, super short
scan, interior and exterior reconstruction. The references\cite{key-27,key-57,key-58}
are reconstruction algorithms in Fourier domain. The references\cite{key-29,key-30,key-32,key-41,key-50,key-52,key-53,key-55,key-63}
are reconstruction algorithms with FBP method. The references\cite{key-36,key-47,key-48,key-69}
are reconstruction algorithms with total variation (TV) minimization
or compressed sensing method. The references\cite{key-35,key-37,key-39,key-51}
are reconstruction algorithms with iterations. The references\cite{key-33,key-46}
are reconstruction algorithms with GPU accelerate. The reference\cite{key-28}
is reconstruction with neural network algorithm. The reference\cite{key-31}
is the reconstruction algorithm with multiresolution. The reference\cite{key-42}
is reconstruction with dual-source. The reference\cite{key-66} introduces
the reconstruction with a Laplace operator. The extrapolation method\cite{key-76,key-77,MattiasNilsson}
and adapted extrapolation method\cite{key-70,key-71,key-74,key-75}
to solve LFOV problem.

\subsection{These author's contributions}

Two generalized iterative refinement methods (IRM), i.e. sub-regional
iterative refinement method (SIRM) and local-region iterative refinement
method (LIRM) for inverse problem or CT image reconstruction with
FFOV are introduced, which can reduce the artifacts and keep the noise
not increased too much. In SIRM the reconstructed image has been divided
to small pieces and re-projected, reconstructed to produce the final
reconstruction image. SIRM was used to overcome the problem of the
TIRM (i.e. the artifacts reduction is at the price of increasing the
noises).

Among NIRM, TIRM, LIRM and SIRM. there are two parameters which are
the margin size of the sub-region and the size of sub-region. The
margin size is originally introduced to eliminate the grids on the
reconstructed image of the SIRM. These authors found that if the margin
size is between 0 and the size of image and the sub-region is chosen
as small as one pixel (the size of sub-region equals 1), the SIRM
becomes LIRM. If the margin size also closes to 0 then the LIRM becomes
TIRM. If the margin size is large enough such as same size of image,
the LIRM becomes NIRM. Hence, the relationship among the SIRM, LIRM,
TIRM and NIRM are summarized. These authors have proved that LIRM
is a special situation of SIRM and proved that NIRM and TIRM are special
situation of LIRM. Hence LIRM and SIRM are two generalized IRM. 

The examples of LIRM in simple inverse problem is introduced to learn
the concept of LIRM, which are also suitable to SIRM. these authors
did not implement LIRM for CT image reconstruction, since it is very
time consuming. A CT reconstruction example using SIRM is done.

\subsection{Arrangement in this article}

In section 2 these authors review the reconstruction method without
iterative refinement (NIRM). In section 3 TIRM is discussed. In section
4 LIRM is discussed . In section 5 the SIRM is discussed.

\section{Local inverse to solve under determinate equation}

Assume the determinate equation is

\[
T\,z=h
\]
where $T$ is a $m\times m$ matrix, $z$ is a unknown vector with
length $m$. $z\in Z$ $h$ is a known vector of length $m$. $h\in H$.
$Z$ and $H$ are definite region of $z$ and $h$. Assume
\[
T=\left[\begin{array}{cc}
A & B\\
C & D
\end{array}\right]
\]
$A$, $B$, $C$, $D$ are submatrix of $T$, 
\[
h=\left[\begin{array}{c}
f\\
g
\end{array}\right]
\]
\[
z=\left[\begin{array}{c}
x\\
y
\end{array}\right]
\]
$f$ and $g$ are subvector of $h$. $x$ and $y$ are subvector of
$z$. We assume $x\in X$, $y\in Y$. Here $X+Y=Z$. We assume $X$
is the interior region or region of interest (ROI). $Y$ is the outside
region or outer side of ROI. We assume $f\in F$, $g\in G$, $G+F=H$.
$F$ is called field of view (FOV). $G$ is called outside of FOV. 

\[
\left[\begin{array}{cc}
A & B\\
C & D
\end{array}\right]\left[\begin{array}{c}
x\\
y
\end{array}\right]=\left[\begin{array}{c}
f\\
g
\end{array}\right]
\]

In the case only $f$ is known, the above equation can be written
as

\[
\left[\begin{array}{cc}
A & B\end{array}\right]\left[\begin{array}{c}
x\\
y
\end{array}\right]=f
\]
This is referred to under determinate equation which has infinite
solution. One of the important solution is minimal norm solution.

\[
\left[\begin{array}{c}
x^{(0)}\\
y^{(0)}
\end{array}\right]=\left[\begin{array}{cc}
A & B\end{array}\right]^{+}f
\]
The subscript ``$^{+}$'' means generalized inverse which gives
the the solution corresponding to minimal norm solution. subscript
``$^{(0)}$'' corresponding to the solution of first iteration. 

The minimal normal solution is not the best solution. The following
we try to further improve the result with so called local inverse
method. 

Assume we are more interest to know ``$x$''. We can subtract the
contribution of $y^{(0)}$from $f$,

\[
f^{(0)}=f-B\,y^{(0)}
\]
We can calculate the solution $x^{(1)}$ the following way,
\[
x^{(1)}=A^{+}f^{(0)}
\]

We found that the above solution has no any improvement, i.e.

\[
x^{(1)}=x^{(0)}
\]
However we can change the formula of $f^{(0)}$ to
\[
f^{(0)}=f-B\,\mathrm{modify}(y^{(0)})
\]
The above function ``$\mathrm{modify}$'' is just modify the mean
value or flop according some priori. In this way we can improve the
result a lot. The following is a example.

The local inverse is introduced with very simple example. Assume

\[
A=\left[\begin{array}{cccc}
3 & 4 & 7 & 6\\
8 & 5 & 8 & 7\\
3 & 6 & 9 & 8\\
4 & 2 & 8 & 9
\end{array}\right]
\]

\[
B=\left[\begin{array}{cccc}
4 & 5 & 6 & 7\\
8 & 7 & 6 & 5\\
6 & 3 & 4 & 6\\
4 & 8 & 4 & 10
\end{array}\right]
\]
\[
x=\left[\begin{array}{c}
3\\
4\\
3\\
6
\end{array}\right]
\]

\[
y=\left[\begin{array}{c}
2\\
6\\
4\\
8
\end{array}\right]
\]
Assume

\[
z=\left[\begin{array}{c}
x\\
y
\end{array}\right]
\]
We can calculate
\[
A\,x+B\,y=f
\]
Now we assume we know vector $f$ and matrix $A$, $B$, we need to
solve $x$ and $y$. We can solve it with least square method so that
can be write as 
\[
z^{(0)}=[A+B]^{+}f
\]

\[
x^{(0)}=first4(z^{(0)})
\]

\[
y^{(0)}=last4(z^{(0)})
\]
We can calculate the errors
\[
error_{x}^{(0)}=\sum|x^{(0)}-x|
\]

\[
error_{y}^{(0)}=\sum|y^{(0)}-y|
\]
\[
f_{x}^{(0)}=f-B\,k_{y}\,y^{(0)}
\]
$modified(y^{0})=k_{y}\,y^{(0)}$
\[
f_{y}^{(0)}=f-B\,k_{x}\,x^{(0)}
\]
$modified(x^{(0)})=k_{x}\,x^{(0)}$
\[
x_{1}=A^{+}f_{x0}
\]

\[
y_{1}=B^{+}f_{y0}
\]

\[
error_{x}^{(1)}=\sum|x^{(1)}-x|
\]

\[
error_{y}^{(1)}=\sum|y^{(1)}-y|
\]

\section{Nor iterative refinement reconstruction method (NIRM)}

Assume the parallel-beam projections are known as $p=p(\theta,u)$,
where $\theta$ is projection angle and $u$ is the index of detector
elements. Assume the projection operator $P$ is known, which is 2D
Radon transform. Assume the non-iteration reconstruction operator
$R$ is also known, which is for example FBP reconstruction. However,
the object or original image $X_{o}(x)$ is unknown. Here $x$ is
the coordinates of the image pixel. The forward equation of the problem
can be write as

\begin{equation}
P\,X=p=p_{o}+p_{n}\label{eq:0-10}
\end{equation}
where $X$ is the unknown image of the above equation. $p_{o}$ is
projections without noises 
\begin{equation}
p_{o}=P\,X_{o}\label{eq:0-10a}
\end{equation}
$p_{n}$ indicates noises in projections. The image can be reconstructed
as

\begin{equation}
X^{(0)}=R\,p\label{eq:0-20}
\end{equation}
$X^{(0)}$ is the first iteration of the reconstructed image using
the reconstruction operator $R$. The superscripts $(0)$ indicates
that $X^{(0)}$ is the first reconstruction in the iteration which
will appear in the following paragraph. The above two formulas can
be written as

\begin{equation}
X^{(0)}=J\,X=J\:X_{o}+R\:p_{n}\label{eq:0-30}
\end{equation}
The first term of the right of the above equation is the signal dependent
reconstructed image. The second term is noise dependent reconstructed
image. In the above equation, $J$ is the projection-reconstruction
operator which is defined as

\begin{equation}
J\equiv R\,P\label{eq:0-40}
\end{equation}
where $x$ is the index of the image pixel before projection-reconstruction
operation. $y$ is the index of pixel of the image after the projection-reconstruction
operation. The operation of Eq.(\ref{eq:0-40}) is defined as

\begin{equation}
J\,X=\sum_{x}j(y,x)\,X(x)\label{eq:0-55}
\end{equation}
Here $j(y,x)$ is the kernel of operator $J$. $J$ can be written
as

\begin{equation}
J=\sum_{x}j(y,x)\,\bullet\label{eq:6-0}
\end{equation}
Where ``$\bullet$'' is multiplication. 

The identical operator is defined in the following
\begin{equation}
I=\sum_{x}\delta(y,x)\,\bullet\label{eq:6-2}
\end{equation}
where

\begin{equation}
\delta(y,x)=\left\{ \begin{array}{c}
1\,\,\,\,\,\,\,\,\textrm{if}\,\,\,\,y=x\\
0\,\,\,\,\,\,\,\,\,\,\,\,\,\,\,else
\end{array}\right.\label{eq:0-50}
\end{equation}
From the above definition there is,

\begin{equation}
\sum_{y}\:\delta(y,x)=1\label{eq:6}
\end{equation}
In the ideal reconstruction the reconstructed image is same as the
original image that means $J\rightarrow I$. ``$\rightarrow$''
is ``close to''. $J$ should satisfy the unitary condition same
as $I$

\begin{equation}
\sum_{y}j(y,x)=1\label{eq:6-1}
\end{equation}

Assume $x'$ is any pixel in the image. The above equation can be
rewritten as

\begin{equation}
\sum_{y}\sum_{x}\:j(y,x)\:\delta(x-x')=1\label{eq:6A}
\end{equation}
The above equation means that the delta function $\delta(x-x')$ after
projection and reconstruction operation and sum operation $\sum_{y}$,
$1$ is obtained.

The unitary condition guarantees that the dc-composition is not changed
after the operation $J=R\:P$. $j(y,x)$ is referred as the resolution
function, $J$ is referred as the resolution operator. In the above
situation it is the resolution function of NIRM. The error of NIRM
is defined as

\begin{equation}
Err^{(0)}\equiv X^{(0)}-X_{o}=(J-I)\:X_{o}+R\:p_{n}\label{eq:0-51}
\end{equation}

The first term is corresponding to artifacts of the method. The second
term is corresponding to noises of the method. Since the resolution
function $J$ is not an exact unitary operator $I$, even if the noises
$p_{n}=0$, it is true according to Eq.(\ref{eq:0-51}) in general
that 
\begin{equation}
X^{(0)}\neq X_{o}\label{eq:0-60}
\end{equation}
One of the problem of the image reconstruction is how to improve the
results of the reconstruction $X^{(0)}$ to let the reconstructed
image more close to the original object function $X_{o}$ ? This can
be rephrased as to find an algorithm with its resolution function
more close to $\delta(y,x)$ function than the resolution function
of NIRM $j(x,y)$ in the same time the noise term does not increase
heavily. The IRMs (TIRM, LIRM and SIRM) deals this problem in the
following sections. These methods have different noise characters
and computation complicity.

\section{Traditional iterative refinement reconstruction method (TIRM)}

In order to further improve the reconstruction results $X^{(0)}$,
TIRM was proposed. TIRM is iterative algorithm with the reconstruction
and re-projection processes. It was obtained in the same way as the
IRM to solve linear equation with an inaccurate known inverse operator.
It was used so many times in different fields and hence it is difficult
to find all of the sources of it. A few examples can be seen \cite{Chang,Zeng,Riddell,OSullivan,Delaney}.
The processes of TIRM is given in the following. The errors i.e. the
differences between the projections $p$ and re-projections $P\,X^{(0)}$
are calculated: $Err=(p-P\,X^{(0)})$. The errors are utilized to
correct the reconstruction $X^{(1)}=X^{(0)}+R\,Err$. The algorithm
can be summarized in the following
\begin{equation}
X^{(1)}=X^{(0)}+R\,(p-P\,X^{(0)})\label{eq:2}
\end{equation}
where $X^{(0)}$ is obtained in Eq.(\ref{eq:0-20}). Substituting
Eq.(\ref{eq:0-20}) to the Eq.(\ref{eq:2}), considering Eq.(\ref{eq:0-40})
the above algorithm can be seen as a filtering algorithm

\begin{equation}
X^{(1)}=F_{TIRM}X^{(0)}\label{eq:3}
\end{equation}
Where the filtering function is defined as

\begin{equation}
F_{TIRM}\equiv2\,I-J\label{eq:4}
\end{equation}
Considering Eq.(\ref{eq:6}, \ref{eq:6-1}) the filtering function
$F_{TIRM}$ satisfies the unitary condition

\begin{equation}
\sum_{y}f_{TIRM}(y,x)=1\label{eq:7}
\end{equation}
$f_{TIRM}(y,x)$ is kernel function of operator $F_{TIRM}$. The results
of TIRM and NIRM will be utilized as contrast to LIRM which will be
discussed in the next paragraph. The resolution operator of TIRM is
$F_{TIRM}\:J$, which is really more close to identical operator $I$
than the resolution operator of NIRM $J$ does. However TIRM is rarely
directly applied to clinical situation since it is sensitive to noises.
This can be seen by further modify the Eq.(\ref{eq:2}) using Eq.(\ref{eq:0-10}),
Eq.(\ref{eq:0-10a}) and Eq.(\ref{eq:0-20}) 

\begin{equation}
X^{(1)}=F_{TIRM}\:J\:X_{o}+F_{TIRM}R\:p_{n}\label{eq:3a}
\end{equation}
Hence the errors

\begin{equation}
Err^{(1)}\equiv X^{(1)}-X_{o}=(F_{TIRM\:}J-I)\:X_{o}+F_{TIRM}R\:p_{n}\label{eq:3aA}
\end{equation}
 The first term of the above formula is a signal dependent image,
which is referred to as artifacts. The second term of the formula
is the noise dependent image. It can be compared the above formula
with Eq.(\ref{eq:0-51}). Since normally $F_{TIRM}\:J$ is closer
to $I$ than $J$, the first term $F_{TIRM}\:J\:X_{o}$ is closer
to $X_{o}$ than $J\:X_{0}$ does. This means that TIRM can get a
more accurate reconstruction corresponding the first term. However
the second term $F_{TIRM}R\:p_{n}$ is normally larger than $R\:p_{n}$.
This means the TIRM is more noise sensitive than NIRM.

Recently a modified TIRM by Johan can be found in reference\cite{JohanSunnegardh}.
In Johan's iterative refinement reconstruction (JIRM), a pre-filtering
process, and two regularization filtering processes are added to TIRM.
Johan has compared the resolution of his method JIRM with NIRM, Johan's
result the Figure 5.7 of \cite{JohanSunnegardh}. a) in Figure 5.7
of \cite{JohanSunnegardh} shows the resolution function of NIRM.
b) in Figure 5.7 of \cite{JohanSunnegardh} shows the resolution function
JIRM without pre-filtering. c) in Figure 5.7 of \cite{JohanSunnegardh}
shows the resolution function of JIRM with pre-filtering. It can be
seen that in b) in Figure 5.7 of \cite{JohanSunnegardh}, there is
a black ring surrounding the white dot. This black ring is called
over-correction. Over-correction increases the resolution but it leads
a noise increase too. Johan used the pre-filtering process to overcome
the problem of over-correction and noises. He first pre-filtered the
projection data and applied the filtered data to his reconstruction.
It is well know that the pre-filtering process does not only smooth
the noise data but also reduced the information contained in the data.
This is why many doctor prefer to see the noise data than the smoothed
data after a filtering process. Pre-filtering process often is not
accepted. The following section will shows the method to overcome
the problem of over-correction in TIRM.

\section{Local-region iterative refinement reconstruction method (LIRM)}

\subsection{The method}

In this section another IRM is defined which is a special case of
next section. First two operators $K$ and $H$ are define as following
\begin{equation}
K\:X=\sum_{x}k(y,x)\:X(x)\label{eq:sec4-01}
\end{equation}
\begin{equation}
H\:X=\sum_{x}h(y,x)\:X(x)\label{eq:sec4-02}
\end{equation}
where

\begin{equation}
k(y,x)=\begin{cases}
\begin{array}{cc}
j(y,x) & \mathrm{if}\:|x-y|\leq r\\
0 & \mathrm{else}
\end{array}\end{cases}\label{eq:sec4-10}
\end{equation}

\begin{equation}
h(y,x)=\begin{cases}
\begin{array}{cc}
0 & \mathrm{if}\:|x-y|<r\\
j(y,x) & \mathrm{else}
\end{array}\end{cases}\label{eq:sec4-20}
\end{equation}
where $r$ is a constant parameter. It defines a region $[y-r,y+r]$
close to $y.$ $r$ is called the margin size. The margin will be
explained in next section. In this section it is assumed that the
image is 1-dimensional. But the result is easy to extent to 2 or 3
dimensional situation. It is clear that there is the relation:

\begin{equation}
J=K+H\label{eq:sec4-30}
\end{equation}
The Local-region iterative refinement method (LIRM) is defined as
following,

\begin{equation}
X^{(1)}=\eta\:R\:p-R\:(P\:\Omega)\:X^{(0)})\label{eq:sec4-40}
\end{equation}
where $\eta$ is a normalized parameter which will be decided later
and the operator $\Omega$ is defined as:

\begin{equation}
\Omega(y,x)=\begin{cases}
\begin{array}{cc}
0 & \mathrm{if}\:|x-y|\leq r\\
1 & \mathrm{else}
\end{array}\end{cases}\label{eq:sec4-50}
\end{equation}

In this article all operators are linear operator except the operator
$\Omega$ which is a normal multiplication operator, i.e.:

\begin{equation}
(J\:\Omega)\:X=\sum_{x}(j(y,x)\:\Omega(y,x))\:X(x)\label{eq:sec4-60}
\end{equation}
Considering Eq.(\ref{eq:0-20}) and Eq.(\ref{eq:0-40}), Eq.(\ref{eq:sec4-40})
can be rewritten as

\begin{equation}
X^{(1)}=\eta\:X^{(0)}-(J\:\Omega)\:X^{(0)}\label{eq:sec4-70}
\end{equation}
considering 
\begin{equation}
J\:\Omega=H\label{eq:sec4-71}
\end{equation}
the LIRM can rewritten as

\begin{equation}
X^{(1)}=\eta\:X^{(0)}-H\:X^{(0)}\label{eq:sec4-80}
\end{equation}
The above formula is rewritten as

\begin{equation}
X^{(1)}=F_{LIRM}X^{(0)}\label{eq:sec4-90}
\end{equation}
where

\begin{equation}
F_{LIRM}=\eta\:I\:-H=\eta\:I-J+K\label{eq:sec40-100}
\end{equation}
$F_{LIRM}$ is the filtering function of the LIRM. Because $J$ can
implemented as $P\:R$, $K$ can be implemented as a small matrix
when $r$ is small. That is why the right part of the above formula
is easier to be implemented than the middle part of the formula. Similar
to Eq.$\:$(\ref{eq:7}) $F_{LIRM}$ should meet the unitary condition:

\begin{equation}
\sum_{y}\sum_{x}f_{LIRM}(y,x)\:\delta(x-x')=1\label{eq:sec4-110}
\end{equation}
This unitary condition means that if the input is $\delta(x-x')$,
$x'$ is any point in the image, the whole output of the filter $F_{LIRM}$
is $1$, since $F_{LIRM}$ should looks like $\delta(y,x)$ function.
The above condition can be write as
\begin{equation}
\sum_{y}f_{LIRM}(y,x)=1\label{eq:sec4-111}
\end{equation}
Considering Eq.$\:$(\ref{eq:sec40-100} and \ref{eq:sec4-111}) implies
that

\begin{equation}
\eta\equiv\eta(x)=1+\sum_{y}h(y,x)=2-\sum_{y}k(y,x)\label{eq:sec4-120}
\end{equation}
The LIRM Eq.(\ref{eq:sec4-40}) can be interpreted as following. The
reconstructed image $X^{(0)}$ is re-projected except the vicinity
of the pixel where the iterative Reconstruction is calculated $(J\:\Omega)\:X^{(0)}$.
The measured projection is reconstructed with $R$, scaled a little
to adjust the dc composition $\eta\:R\:p$. The difference of the
above two projections are used to produced a reconstruction $(\eta\:R\:p-(J\:\Omega)\:X^{(0)})$.
Inside of the vicinity, only the center point is kept, which is the
pixel where the iterative method is done. Other pixel can be obtained
in the same way. The LIRM always concerns in a local region, which
is the vicinity of the pixel. Hence it is referred as the local region
iterative refinement method.

\subsection{In the limit case}

In this article the discussed IRM are generalized inverse methods.
CT image reconstruction can be seen as a example. It is easy to understand
the principle of method to consider the object $X_{o}$ as one dimensional
image. Assume $x\in[-\frac{L}{2},\frac{L}{2}]$ is the coordinates
of the image pixel. Here $L$ is the number of pixels of the image.
if $r\rightarrow L$, then according to Eq.(\ref{eq:sec4-10}, \ref{eq:sec4-20},
\ref{eq:sec4-30} and \ref{eq:sec4-120}) there are

\begin{equation}
K\rightarrow J,\quad H\rightarrow0,\quad\eta\rightarrow1\label{eq:sec4-130}
\end{equation}
Considering the above formula and Eq.(\ref{eq:sec40-100}) there is
\begin{equation}
F_{LIRM}=\eta\:I\:-H\rightarrow I=F_{NIRM}\label{eq:sec4-131}
\end{equation}
i.e. in this case there is LIRM $\rightarrow$NIRM.

if $r\rightarrow0$, the similarly according to Eq.(\ref{eq:sec4-10},
\ref{eq:sec4-20}, \ref{eq:sec4-30}) there are

\begin{equation}
k(y,x)\rightarrow j(y,x)\delta(y,x)\label{eq:sec4-140}
\end{equation}
That is

\begin{equation}
K\rightarrow j(y,y)\:I\label{eq:sec4-141-0}
\end{equation}
considering Eq.(\ref{eq:sec4-120}), there is
\begin{equation}
\eta(x)\rightarrow2-j(x,x)\label{eq:sec4-141}
\end{equation}
and
\begin{equation}
\eta\:I-J+K\rightarrow(2-j(y,y))\:I-J+j(y,y)\:I=2\:I-J\label{eq:sec4-150}
\end{equation}
Considering the above formula and Eq.(\ref{eq:sec40-100}) and Eq.(\ref{eq:4}),
there is 
\begin{equation}
F_{LIRM}\rightarrow F_{TIRM}\label{eq:sec4-151}
\end{equation}
i.e. in this case LIRM $\rightarrow$TIRM. This means that the NIRM
and TIRM are two special cases of LIRM as $r\rightarrow L$ or $r\rightarrow0$.

\subsection{A simple example}

In following example, the original image is 1-dimensional for simplicity.
CT reconstruction is not taken which is at least two dimension. Assume
the forward operator is $P=T\:V$. Here $T$ is assumed as discrete
Fourier transform. The inverse operator is taken as $R=T^{-1}$ ,
$T^{-1}$ is the inverse discrete Fourier transform. Assume the size
of the image is $L=128$. The blurring operator $V$ is assumed as 

\begin{equation}
V(x)=\frac{1}{[(\frac{x}{\alpha L})^{2}+\beta]^{\rho}}\label{eq:sec4-190}
\end{equation}
where $\alpha=0.01$, $\beta=0.2$, $\rho=0.9$. In this example the
resolution function

\begin{equation}
J=R\:P=T^{-1}\:T\:V=V\label{eq:sec4-200}
\end{equation}
In this example $J(x,y)=J(x-y)=V(x-y)$. $F_{NIRM}$, $F_{TIRM}$
and $F_{LIRM}$ are summarized as $F_{a}$. Corresponding 3 IRMs can
be written as following,

\begin{equation}
X^{(1)}=F_{a}\:J\:X_{o}+F_{a}R\:p_{n}\label{eq:sec4-210}
\end{equation}
The error function can be written as,

\begin{equation}
Err_{a}^{(1)}\equiv X^{(1)}-X_{o}=(F_{a}\:J-I)\:X_{o}+F_{a}R\:p_{n}\label{eq:sec4-220}
\end{equation}
Here $a$ is corresponding to NIRM, TIRM and LIRM, and $F_{a}$ has
3 format, $F_{NIRM}=I$, $F_{TIRM}=2\:I-J$, $F_{LIRM}=\eta\:I-J+K$
The noises and artifacts depends on the shape of the operator $F_{a}$

\subsection{Advantages and disadvantages}

In general, the errors of LIRM are less than TIRM meaning that

\begin{equation}
||Err_{LIRM}||\ll||Err_{TIRM}||\label{eq:sec4-180}
\end{equation}
The noises of LIRM are less than TIRM meaning that

\begin{equation}
\sigma_{y}^{2}\{Err_{LIRM}\}\ll\sigma_{y}^{2}\{Err_{TIRM}\}\label{eq:sec4-181}
\end{equation}
Here $\sigma_{y}^{2}$ is defined local variance as

\begin{equation}
\sigma_{y}^{2}\{X(x)\}=\frac{1}{N-1}\sum_{x\in\omega_{y}}(X(x)-E(X(x)))^{2}\label{eq:4-182}
\end{equation}
Where $\omega_{y}$ is the local region close to $y$. $E$ is the
local mean defined as 
\begin{equation}
E_{y}\{X(x)\}=\frac{1}{N}\sum_{x\in\omega_{y}}X(x)\label{eq:4-183}
\end{equation}
Where $N$ is number of pixel(voxel) inside the local region $\omega_{y}$.
$N$ can be found experimentally. these authors are not going to prove
the above formula Eq(\ref{eq:sec4-180}, \ref{eq:sec4-181}), instead
an example in following section will show the results.

\begin{figure}
 \centering

\subfigure[]{

\includegraphics[width=0.5\textheight]{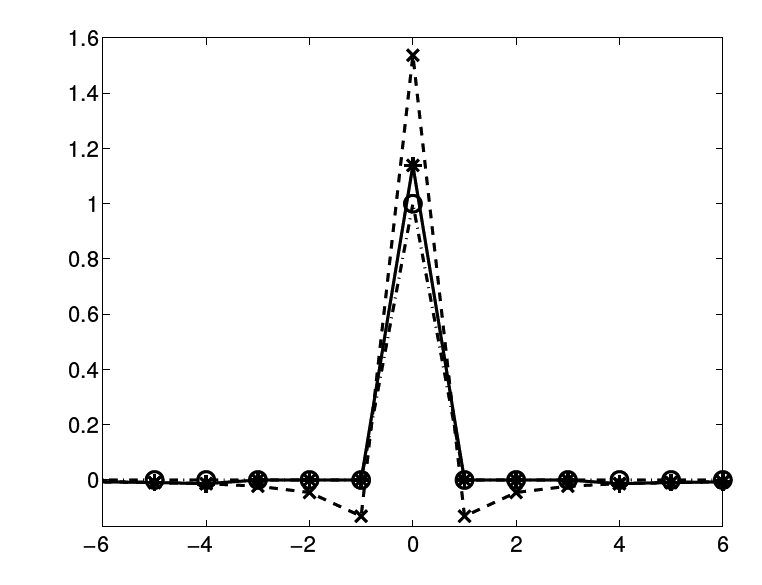}\label{fig:fig01_a}

} \subfigure[]{

\includegraphics[width=0.5\textheight]{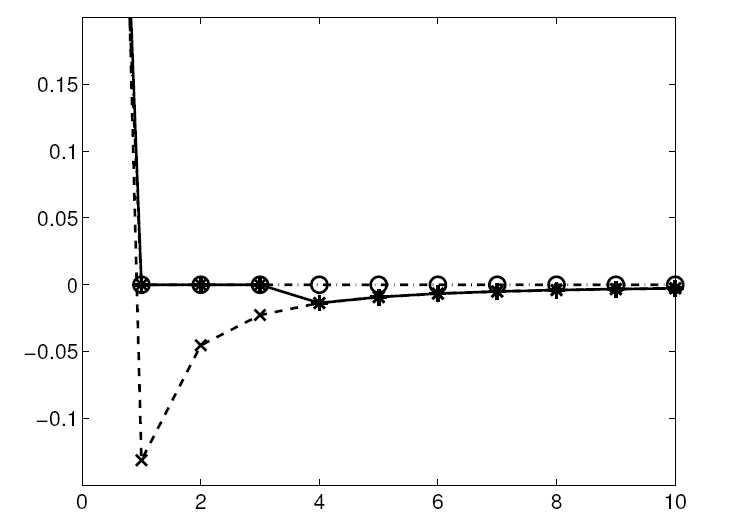}\label{fig:fig01_b}

} 

\caption{(a) The dash-dotted line with circle mark shows $F_{NIRM}(x)$, The
dashed line with cross mark shows $F_{TIRM}(x)$, the solid line with
star mark shows $F_{LIRM}(x)$. (b) is the zoomed image of (a) to
see the vicinity at x=r. Here r=3. It is clear that there is a large
negative values for $F_{TIRM}$ at $r=\pm1.$ }

\label{fig:fig01}
\end{figure}

Figure \ref{fig:fig01} shows the filtering function of $F_{a}$ for
different methods. I is can be seen that $F_{TIRM}$ has a large negative
values at $|x|<r$. This is corresponding the black ring of the resolution
function in b) of Figure 5.7 of the reference\cite{JohanSunnegardh}.
This is also referred as over-correction. The over-correction can
increase the resolution yet it also increases the noises. On the other
hand, the filtering function $F_{LIRM}$ has no large negative value
at $|x|<r$. When $|x|>r$ $F_{LIRM}\simeq F_{TIRM}$ , hence $F_{LIRM}$
can also reduce the artifacts for example beam harden artifacts in
CT reconstruction. In Figure \ref{fig:fig01} these authors choose
$r=3$. $r$ is a parameter can be adjusted according to the size
of the image. Usually the large the image size, the large the $r$
should be. From Figure \ref{fig:fig01} it can be seen that if $r\rightarrow0$,
there is $F_{LIRM}\rightarrow F_{TIRM}$. When $r\rightarrow L$,
there is $F_{LIRM}\rightarrow F_{NIRM}$. 

In the following it is assumed that $X_{o}(x)=\mathrm{sign}(x)$.
The image edge is at the place $x=0$. It is also assumed that the
size of image is $L=2048$, and $r=122$. Here these authors have
increased the size of image to show the results more clearly. Measured
data is simulated with $p_{o}=P\:X_{o}$. Matlab is used to created
the simulated noise: $p_{n}(x)=0.004*randint(1,${[}-5,5{]}). Noise
$p_{n}$ is added to data $p=p_{o}+p_{n}$. Assume the forward operator
is $P=T\:V$ . Here $T$ is assumed as discrete Fourier transform.
The inverse operator is taken as $R=T^{-1}$ , $T^{-1}$ is the inverse
discrete Fourier transform. $V$ is defined in Eq.(\ref{eq:sec4-190}).
The three reconstruction results using the methods NIRM, TIRM, LIRM
are compared.

\begin{figure}
 \centering

\subfigure[]{

\includegraphics[width=0.35\textheight]{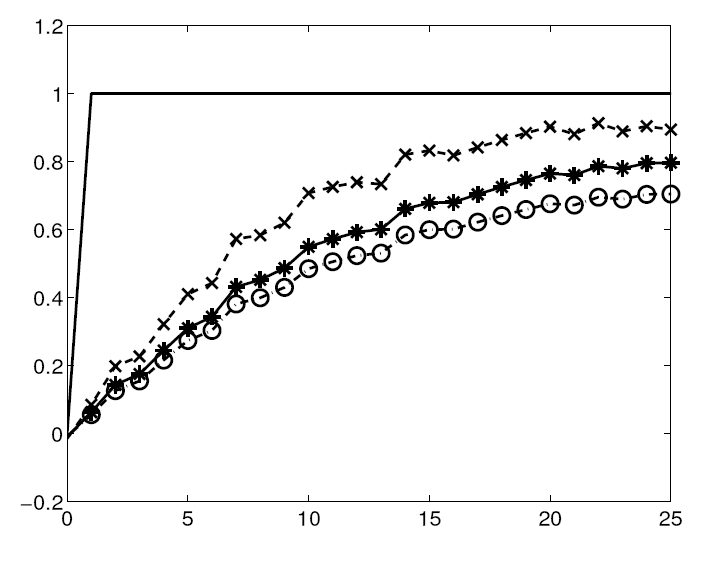}\label{fig:fig02-a}

}\subfigure[]{

\includegraphics[width=0.35\textheight]{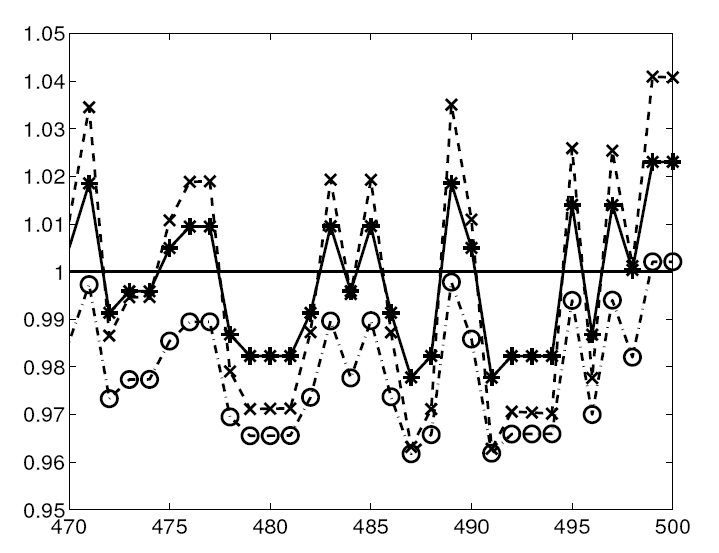}\label{fig:fig02-b}

}

\subfigure[]{

\includegraphics[width=0.35\textheight]{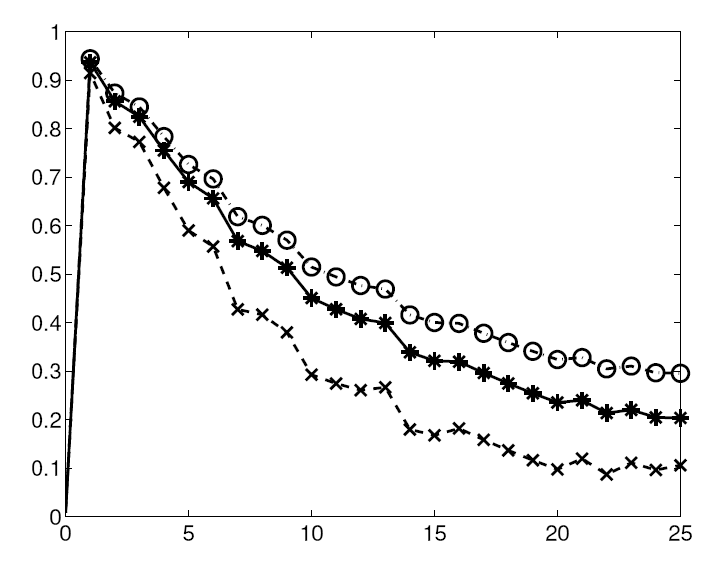}\label{fig:fig02-c}

}\subfigure[]{

\includegraphics[width=0.35\textheight]{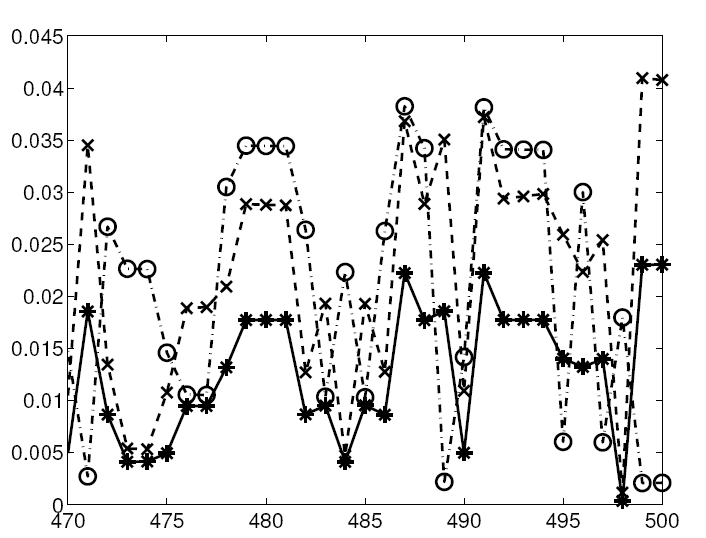}

}\label{fig:fig02-d}

\caption{(a) Reconstructed images of 3 methods at close to the image edges.
(b) indicates the reconstructed images of 3 methods at the places
far away from the image edges. (c) indicates the absolute errors of
the three methods at the place close to the image edges. (d) indicates
the absolute errors of 3 methods at the places far away from the image
edges. The solid line in (a) and (b) are original object function
$X_{o}$. The dashed lines with circle mark corresponding to NIRM.
The dashed line with cross mark corresponding to TIRM. The solid line
with star mark corresponds to LIRM.}

\label{fig:fig02}
\end{figure}

In Figure \ref{fig:fig02} it can be seen that in the place of image
edges the TIRM has the smallest error. In the place far away from
the edges, The LIRM has lowest errors. Parameter $r$ can be adjusted
so the image is optimal at reducing the noise and increasing of the
accuracy. The errors of LIRM are less than NIRM in both place of image
edges and the place far away from the edges. The errors of TIRM are
less than NIRM in the place of image edges but it is at the same level
with NIRM at the place far away from the image edges. Normally it
is acceptable that there are errors at image edges, but the errors
at the place far away from image edges should be as small as possible.
Thus LIRM is better than TIRM and NIRM in decrease errors and artifacts.

It can be seen that the results are not dependent on the operator
of $T$. It is dependent only with $J$. Here it is assumed that $T$
is Fourier Transform to make things easy. Actually if $J=R\:P=V$
and $V\neq I$ (here $I$ is identical operator), the results of all
IRMs (TIRM, LIRM) are not related to the operator $R$ and $P$, but
it is dependent to there product $J=V=R\:P$. 

For example if a image $G$ is filtered with convolution by operator
$F$. Assume the filtered image $G'=F\star G$ is known, ``$\star$''
means convolution. The original image $G$ is required to be recovered
from $G'$ and the known operator $F$. Assume that the Fourier transform
of $F$ is known which is $\tilde{F}=\mathcal{F}\{F$\}, $\mathcal{F}$
is Fourier transform. The recover operator $Q=\mathcal{F^{\mathrm{-1}}}(\frac{1}{\tilde{F}})$
can be defined. $\mathcal{F^{\mathrm{-1}}}$ is inverse Fourier transform.
$Q$ can fully recover the original image, since if $\tilde{Q}=\mathcal{F}\{Q\}$,
$\tilde{Q}\:\tilde{G'}=\tilde{Q}\:\tilde{F}\:\tilde{G}=\frac{1}{\tilde{F}}\:\tilde{F}\:\tilde{G}=\tilde{G}$,
and hence $Q\star G'=G$. However if $\tilde{F}$ has $0$ or very
close to $0$ some where. $\tilde{Q}=\frac{1}{\tilde{F}}$ will have
``$\frac{1}{0}$''. In this situation, the above image recovery
method can not be implemented. Thus a regularization is required.
For example $Q=\mathcal{F}(\frac{1}{\tilde{F}+\alpha})$ can be defined,
here $\alpha$ small number which is a regularization factor. In this
case the recovered image $Q\star G'=\mathcal{F}(\frac{\tilde{F}}{\tilde{F}+\alpha})\star G$.
$J=Q\star F=\mathcal{F}(\frac{\tilde{F}}{\tilde{F}+\alpha})$. $V=\mathcal{F}(\frac{\tilde{F}}{\tilde{F}+\alpha})$
can be referred as the blurring function. The recovery method can
be improved by the IRMs (NIRM, TIRM and LIRM). The results of IRMs
are only related to the resolution function $J=V$. If $V$ is same
as Eq.(\ref{eq:sec4-190}), the recovered image $G'$ will has the
same results as the Fig.$\:$\ref{fig:fig01} and \ref{fig:fig02}. 

Since to implement the LIRM with the example of CT image reconstruction
is very time-consuming, these authors only study the simple examples
in this section which is 1-D inverse problem. Note that although in
this section, the example of image reconstruction has not been done,
the analysis results are still suitable to the situation of CT image
reconstruction. In the next section these authors will study another
IRM, i.e. SIRM, which is close to LIRM, but is easier to implement
for CT image reconstruction.

\section{Sub-regional iterative refinement method (SIRM)}

\subsection{History of SIRM}

\begin{figure}
 \centering

\subfigure[]{

\includegraphics[width=0.29\textheight]{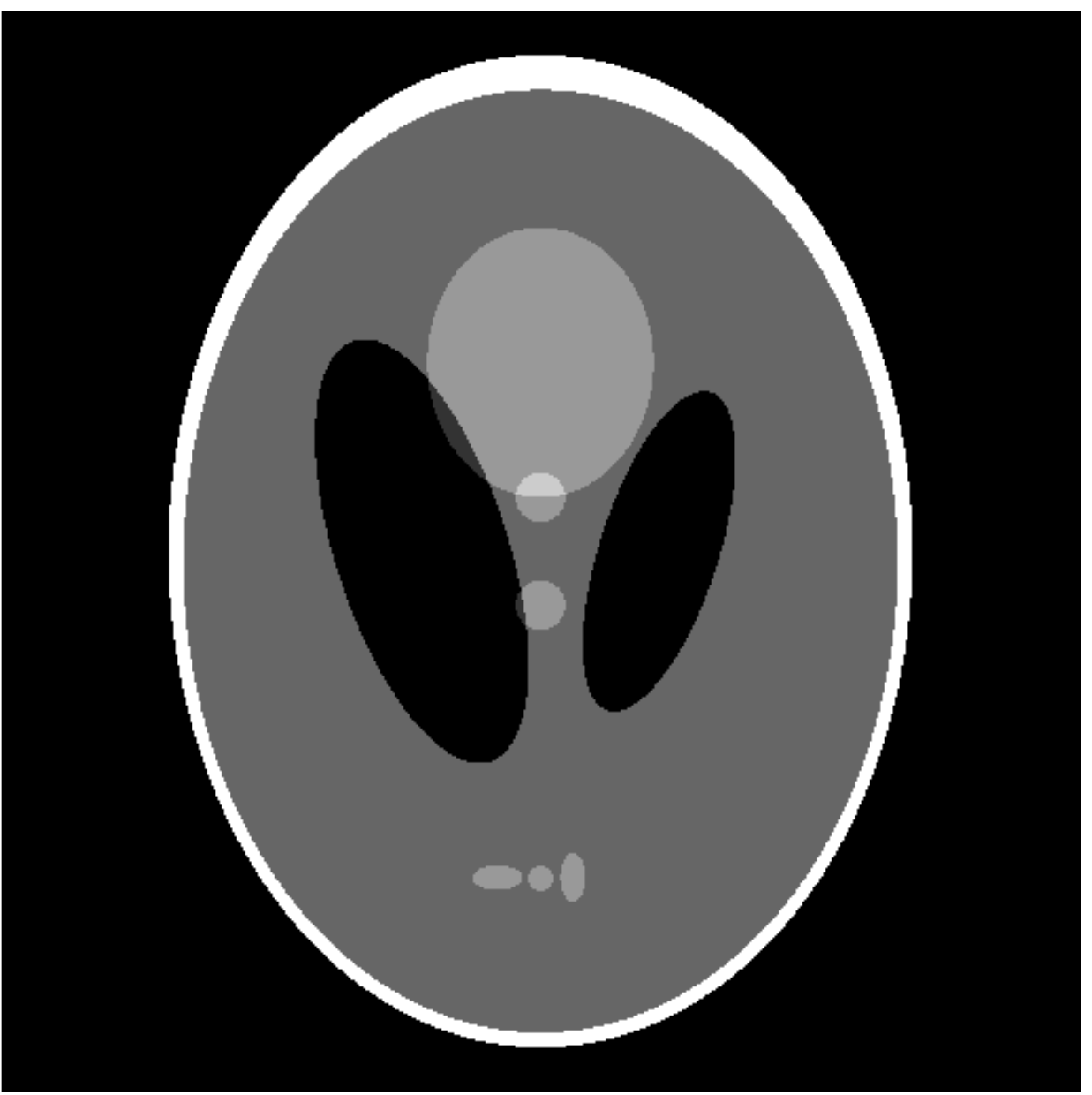}\label{fig:fig4_A}

} \subfigure[]{

\includegraphics[width=0.29\textheight]{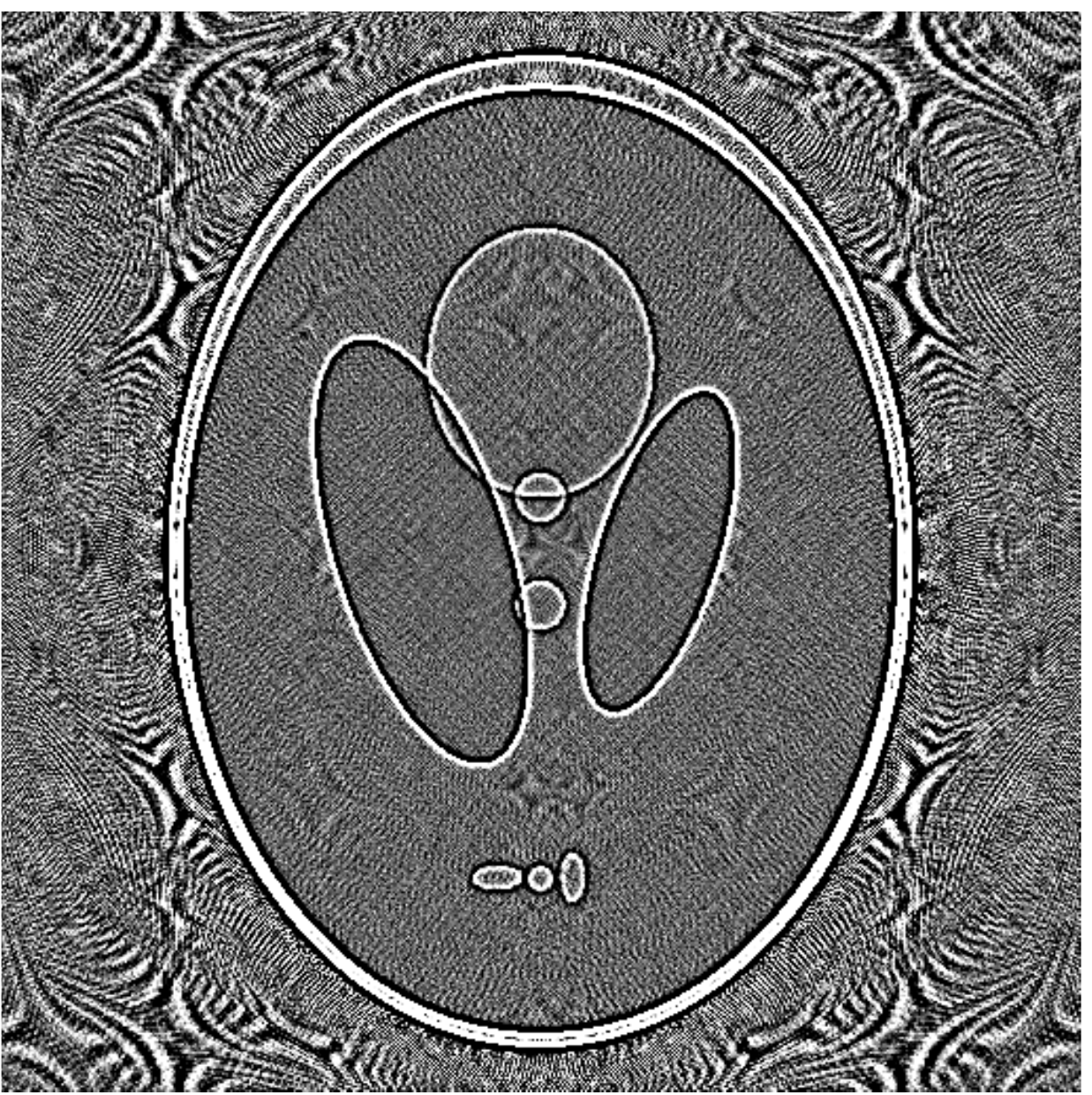}\label{fig:fig4_B}

}

\caption{(a) The phantom. (b) The absolute errors of the FBP reconstruction. }

\label{fig:fig4-new}
\end{figure}

During the work of iterative reconstruction for LFOV\cite{Ref-22-Shuangren-Zhao,Ref-23-ShuangrenZhao},
these authors have noticed that not only the truncation artifacts
are reduced but the normal artifacts are also reduced. Here the normal
artifacts means the artifacts appeared in the reconstruction of FFOV
(full field of view) instead of LFOV (limited field of view). Figure
\ref{fig:fig4-new} shows the normal artifacts appearing with the
FBP method, this results is from Matlab. 

If the local inverse reconstruction for LFOV \cite{Ref-22-Shuangren-Zhao,Ref-23-ShuangrenZhao}
is applied in the situation of FFOV , the algorithm can be summarized
as the following,

\begin{eqnarray}
p^{(1)} & = & p-P\,\Omega\,R\,p\label{eq:5-3a}\\
X^{r} & = & R\,p^{(1)}\label{eq:5-3}
\end{eqnarray}
 Where $p=P\,X$. $\Omega$ is the reverse truncation operator, which
is defined in the following,
\begin{equation}
\Omega(x)=\left\{ \begin{array}{cc}
0 & \textrm{if}\;x\in\textrm{ROI}\\
1 & \textrm{if}\;x\notin\textrm{ROI}
\end{array}\right.\label{eq:5-4}
\end{equation}
Here ROI is the region of interest that is any small arbitrary sub-region
where a reconstruction can be made, and $x$ is the coordinates of
the pixel of the reconstructed image. In the following example it
is assumed that the ROI of the object is a centric disk-shape region
and its radius is half of the radius of the image. Here the disk-shape
region is chosen because these authors start this kind reconstruction
from LFOV which require a disk-shaped region.

Eq.(\ref{eq:5-3}) is a simplification of the algorithm\cite{Ref-22-Shuangren-Zhao}.
The extrapolation process is take away because this is FFOV instead
LFOV. There is no truncation and extrapolation is not necessary. In
order to simplify the algorithm Eq.(\ref{eq:5-3a},\ref{eq:5-3})
further, the truncation operator can be defined as:

\begin{equation}
T(x)=\left\{ \begin{array}{cc}
1 & \textrm{if}\;x\in\textrm{ROI}\\
0 & \textrm{if}\;x\notin\textrm{ROI}
\end{array}\right.\label{eq:5-5}
\end{equation}
The relation between the truncation and the reverse truncation operator
is given in the following

\begin{equation}
\Omega(x)=1(x)-T(x)\label{eq:5-6}
\end{equation}
Here $1(x)$ is unit operator with its value as $1$ everywhere on
the image. Considering the above Eq.(\ref{eq:5-6}), Eq.(\ref{eq:5-3a},\ref{eq:5-3})
can be rewritten as
\begin{eqnarray}
p^{(1)} & = & P\,T\,R\,p+(p-P\,R\,p)\label{eq:5-7a}\\
X^{r} & = & T\,R\,p^{(1)}\label{eq:5-7}
\end{eqnarray}
In order further improved the iterative reconstruction, a truncation
operator $T$ is added to the reconstructed image of the second formula
of Eq.(\ref{eq:5-7}). The truncation operator set zeros outside the
ROI. This helps to delete unwanted image outside the ROI. 

\begin{figure}
 \centering

\subfigure[]{

\includegraphics[width=0.2\textheight]{fig_5A}\label{fig:fig2_a}

} \subfigure[]{

\includegraphics[width=0.2\textheight]{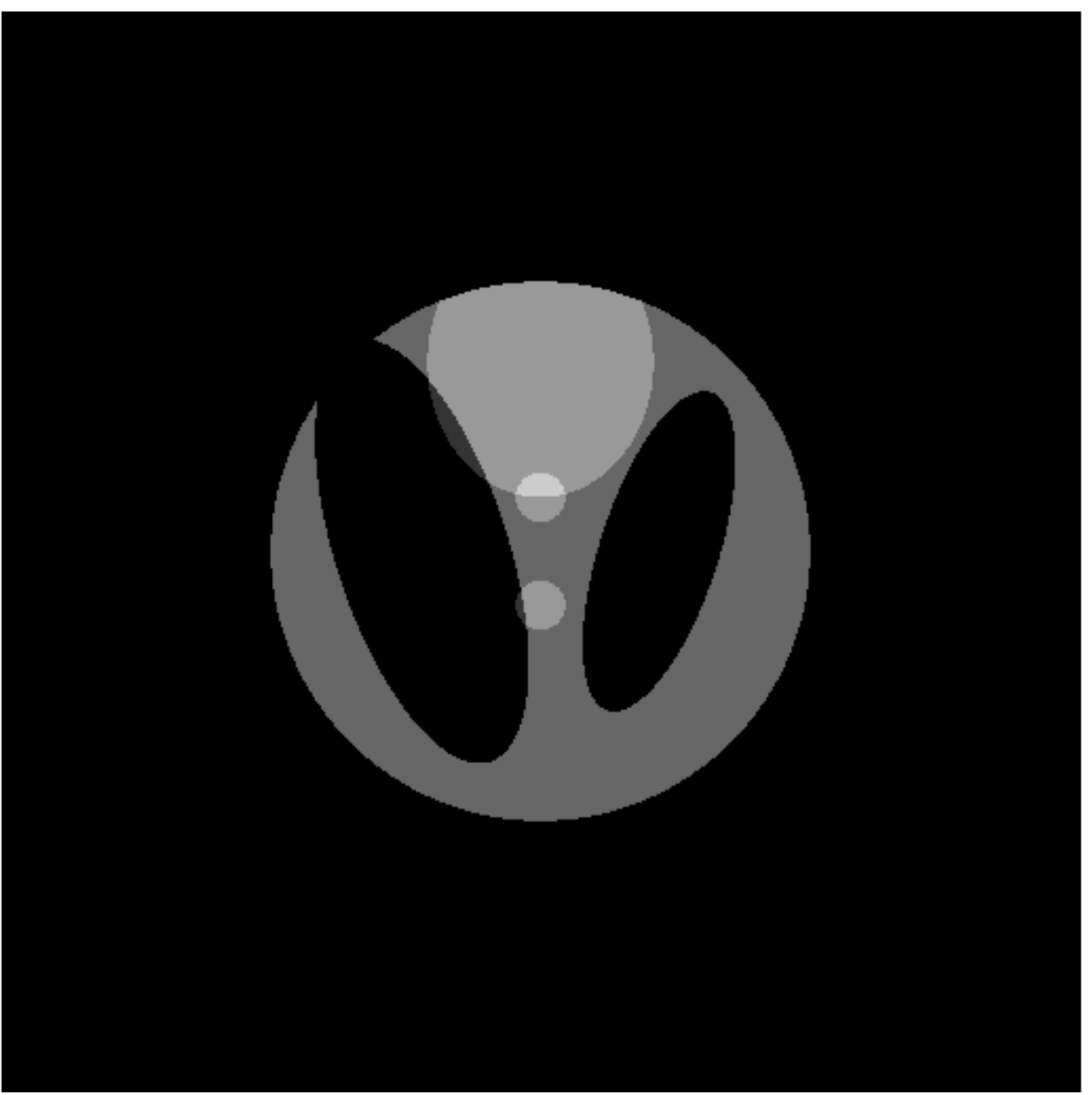}\label{fig:fig2_b}

} \subfigure[]{

\includegraphics[width=0.2\textheight]{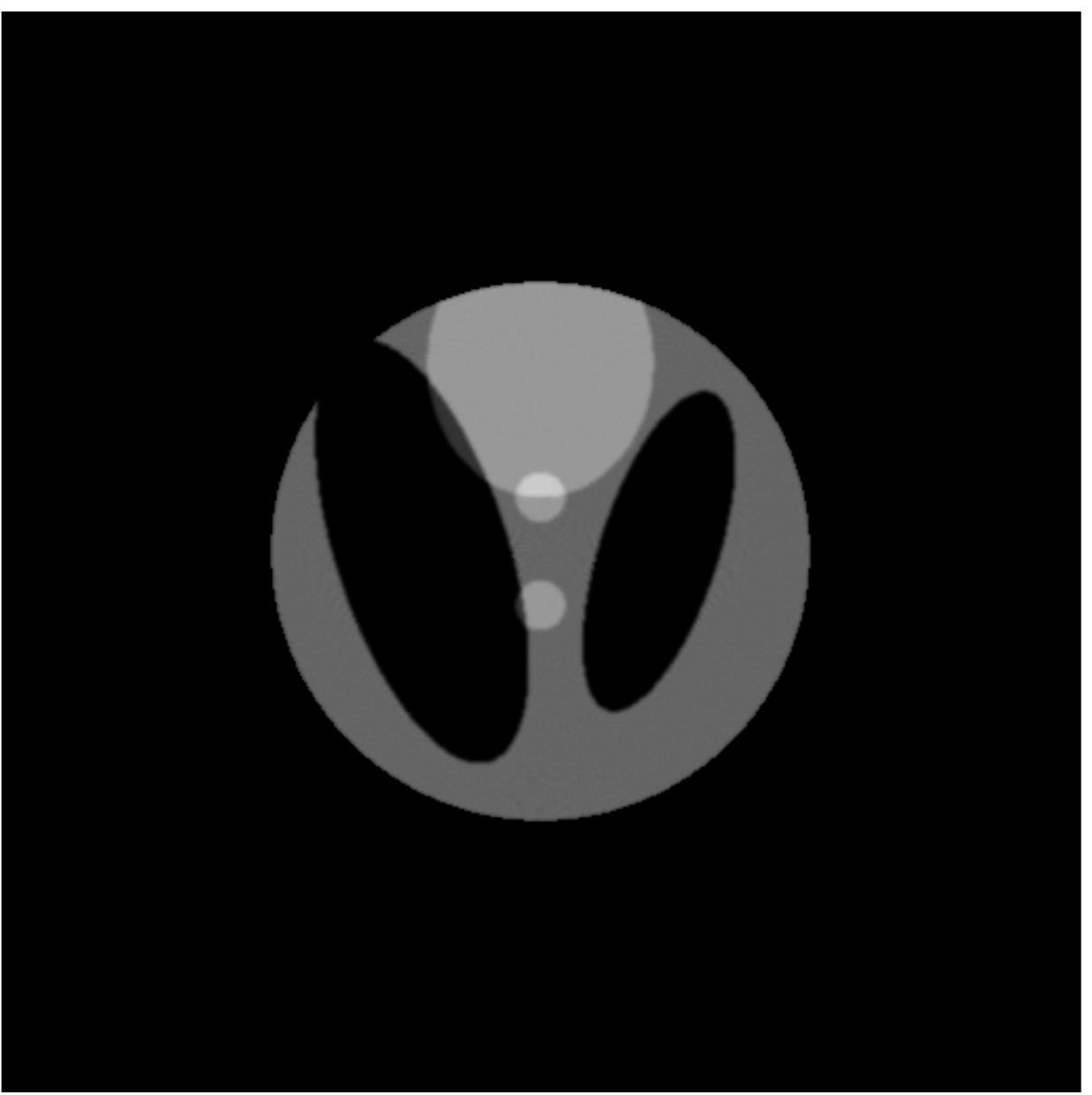}\label{fig:fig2_c}

} \subfigure[]{

\includegraphics[width=0.2\textheight]{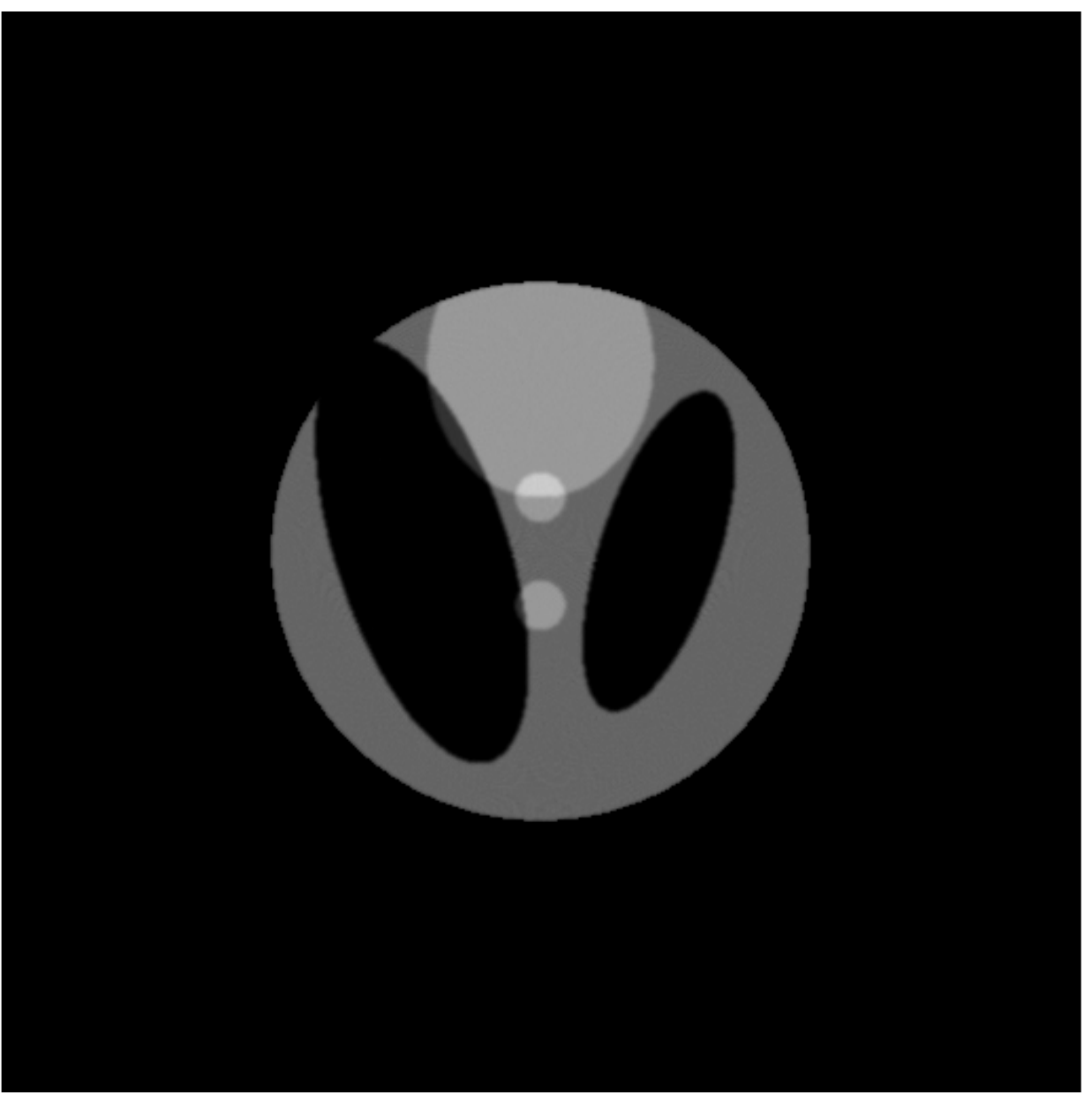}\label{fig:fig2_d}

} \subfigure[]{

\includegraphics[width=0.2\textheight]{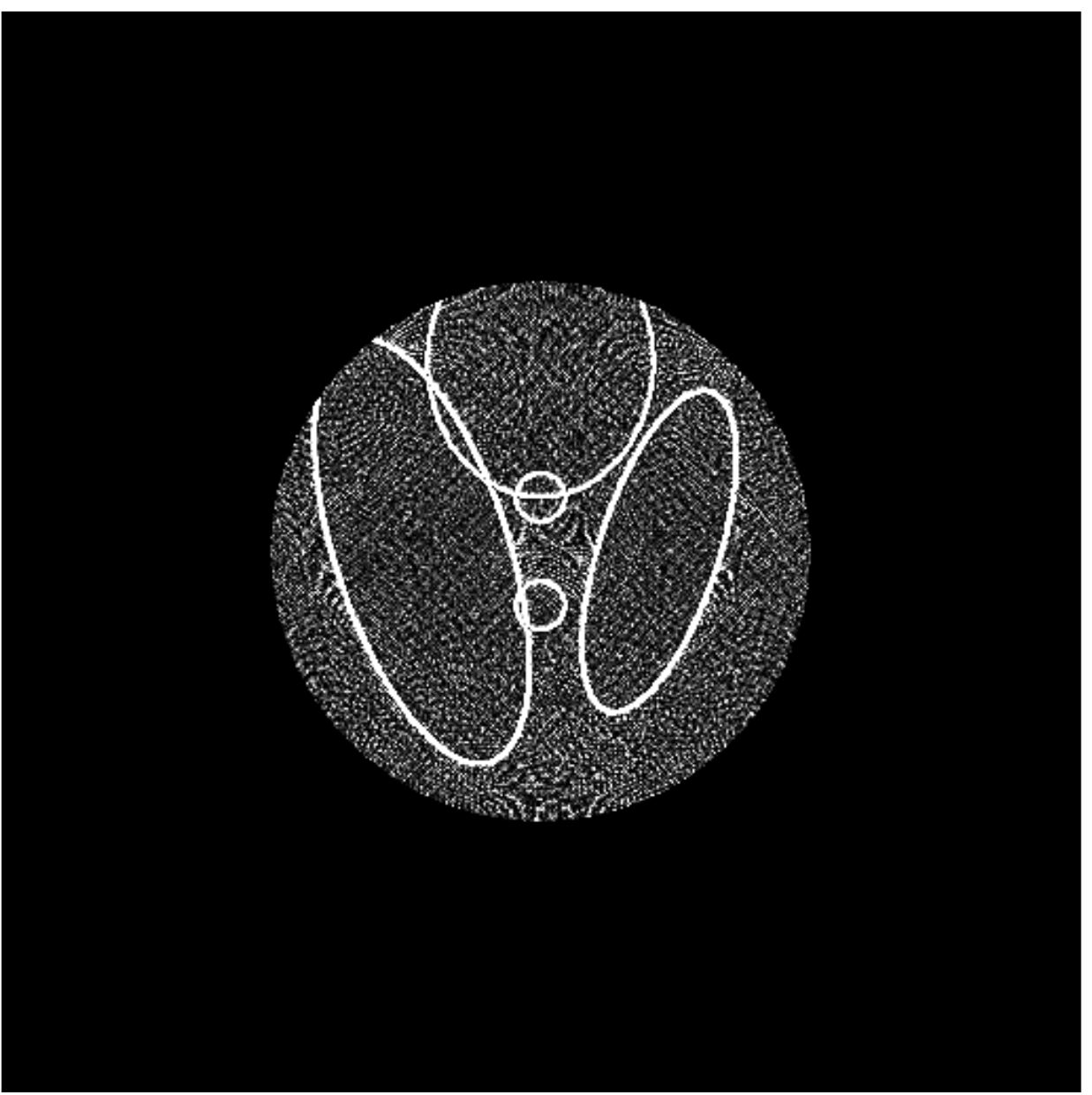}\label{fig:fig2_e}

} \subfigure[]{

\includegraphics[width=0.2\textheight]{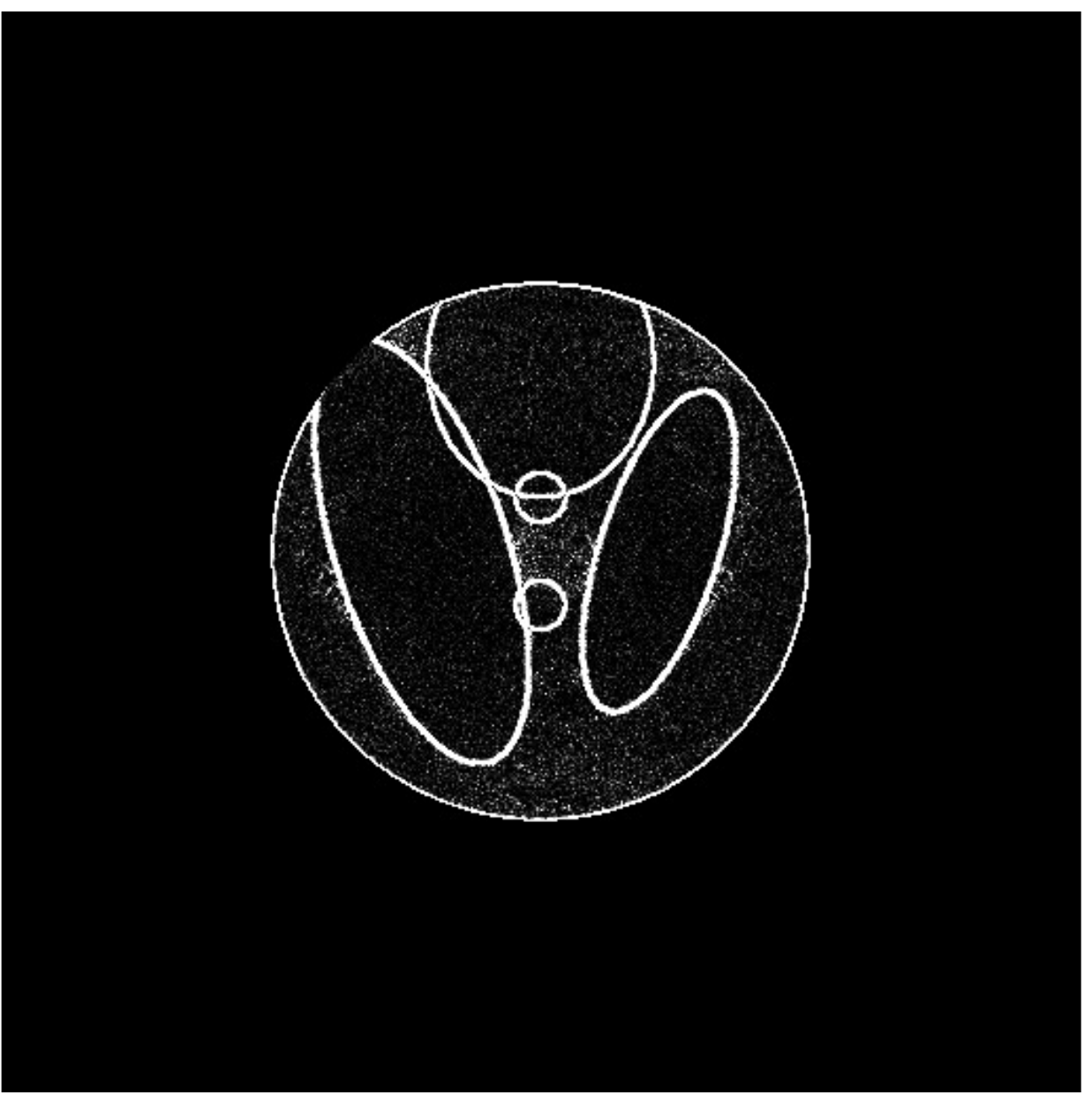}\label{fig:fig2_f}

}

\caption{(a) is image of the Shepp-Logan head phantom. (b) is the crop of the
image of phantom corresponding to the ROI. (c) is the reconstruction
with FBP algorithm. (d) is the reconstruction with the iterative reconstruction
algorithm Eq.(\ref{eq:5-7a}, \ref{eq:5-7}). (e) is the image of
the errors for the FBP method or NIRM. (f) is the image of errors
for the iterative reconstruction of Eq.(\ref{eq:5-7a}, \ref{eq:5-7}).}

\label{fig:fig2}
\end{figure}

Figure~\ref{fig:fig2} offers the reconstruction results for FBP
algorithm and the above iterative algorithm. Figure~\ref{fig:fig2_a}
is the image of the Shepp-Logan head phantom. Figure~\ref{fig:fig2_b}
is the crop of the image of the phantom corresponding to the ROI which
is a centered disk. Figure~\ref{fig:fig2_c} is the reconstruction
with FBP algorithm from the simulated parallel beam projections obtained
from Matlab. The number of projections is 360 for the half circle
scan (180 degree). The space between the two elements of the detector
is taken as the same as the space between the two pixels of image.
The data size of image of phantom is $512\times512$. Figure~\ref{fig:fig2_d}
is the reconstruction with iterative algorithm of Eq.(\ref{eq:5-7a},
\ref{eq:5-7}). The projections and all parameters are same as Figure~\ref{fig:fig2_c}.
It is difficult to see the differences between Figure~\ref{fig:fig2_c}
and Figure~\ref{fig:fig2_d} if you do not see them carefully. An
error function is defined as following 

\begin{equation}
Err=T\,|X^{r}-X|^{2}\label{eq:9}
\end{equation}

Figure~\ref{fig:fig2_e} and Figure~\ref{fig:fig2_f} are error
functions corresponding to Figure~\ref{fig:fig2_c} and Figure~\ref{fig:fig2_d}.
Figure~\ref{fig:fig2_e} and Figure~\ref{fig:fig2_f} use the same
scale of brightness. It is clear that Figure~\ref{fig:fig2_e} is
bright than Figure~\ref{fig:fig2_f} meaning that the reconstruction
errors of FBP algorithm is larger than the iterative algorithm of
Eq.(\ref{eq:5-7a}, \ref{eq:5-7}). It is can be seen that the values
of reconstruction with FBP algorithm are little bit lower than the
values of the phantom. However the values with iterative reconstruction
of Eq.(\ref{eq:5-7a}, \ref{eq:5-7}) are much close to the values
of the phantom. This also shows the improvement of the iterative algorithm
Eq.(\ref{eq:5-7a}, \ref{eq:5-7}).

\begin{figure}
 \centering

\subfigure[]{

\includegraphics[width=0.2\textheight]{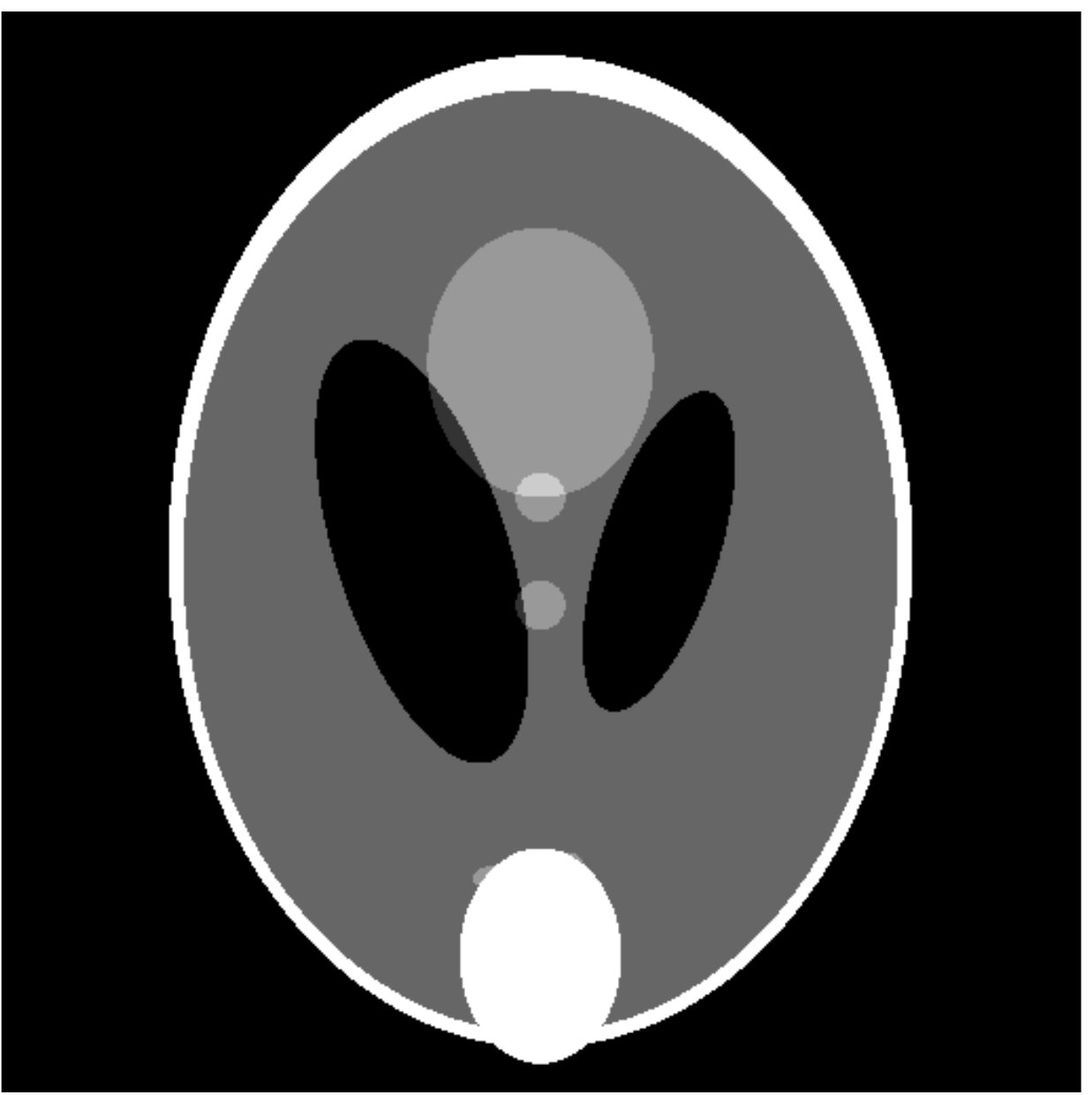}\label{fig:fig4_a}

} \subfigure[]{

\includegraphics[width=0.2\textheight]{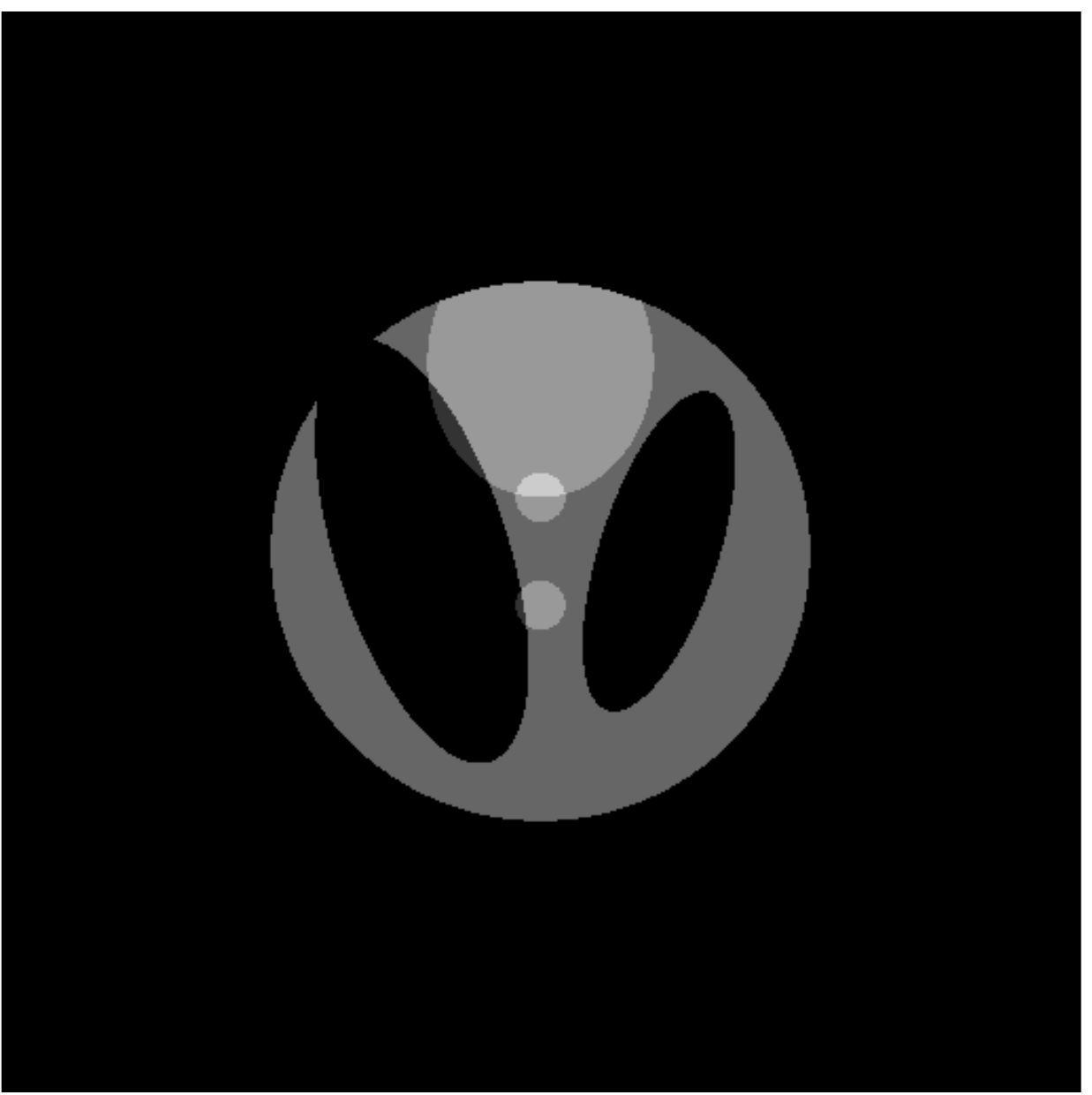}\label{fig:fig4_b}

} \subfigure[]{

\includegraphics[width=0.2\textheight]{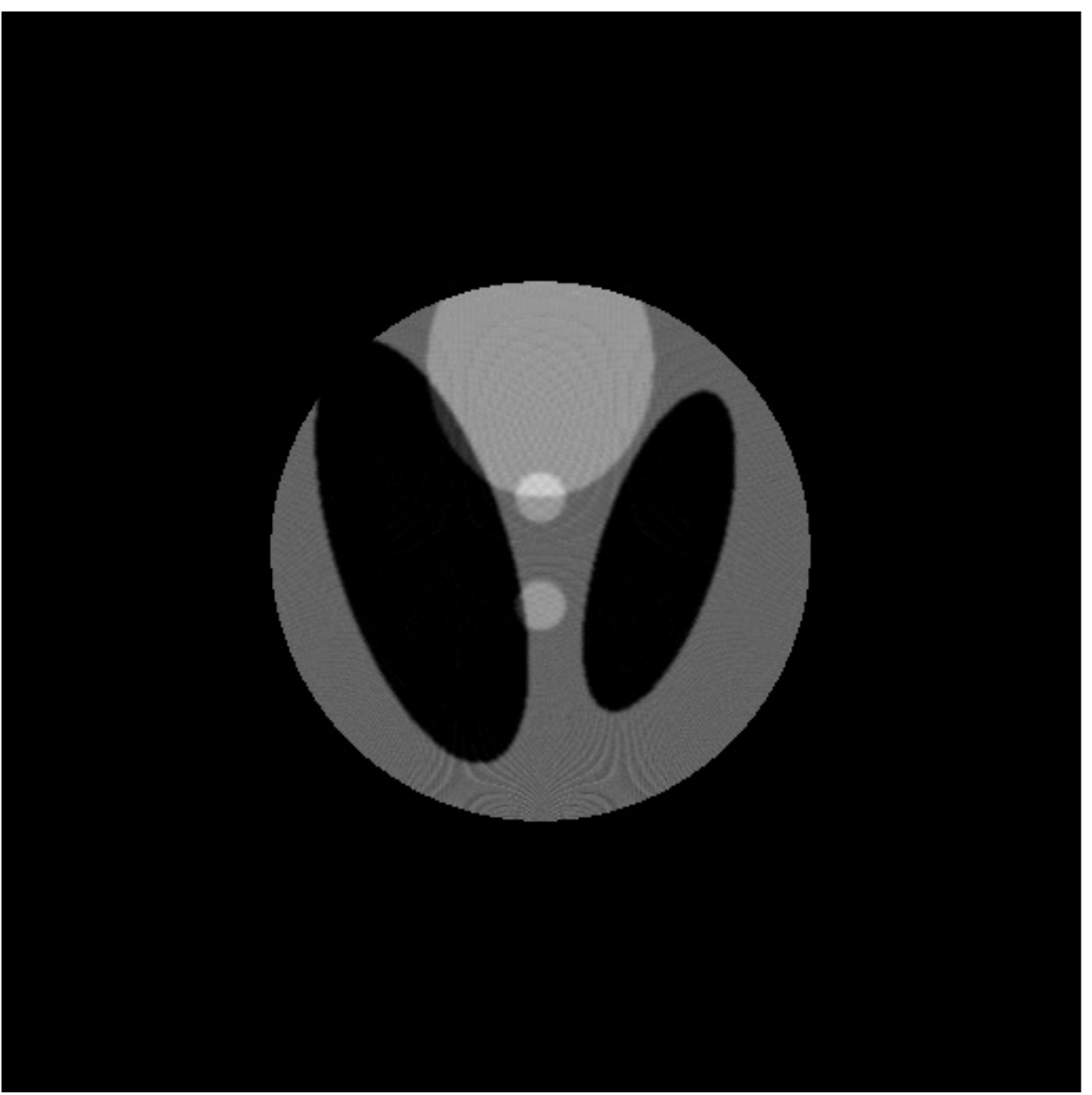}\label{fig:fig4_c}

} \subfigure[]{

\includegraphics[width=0.2\textheight]{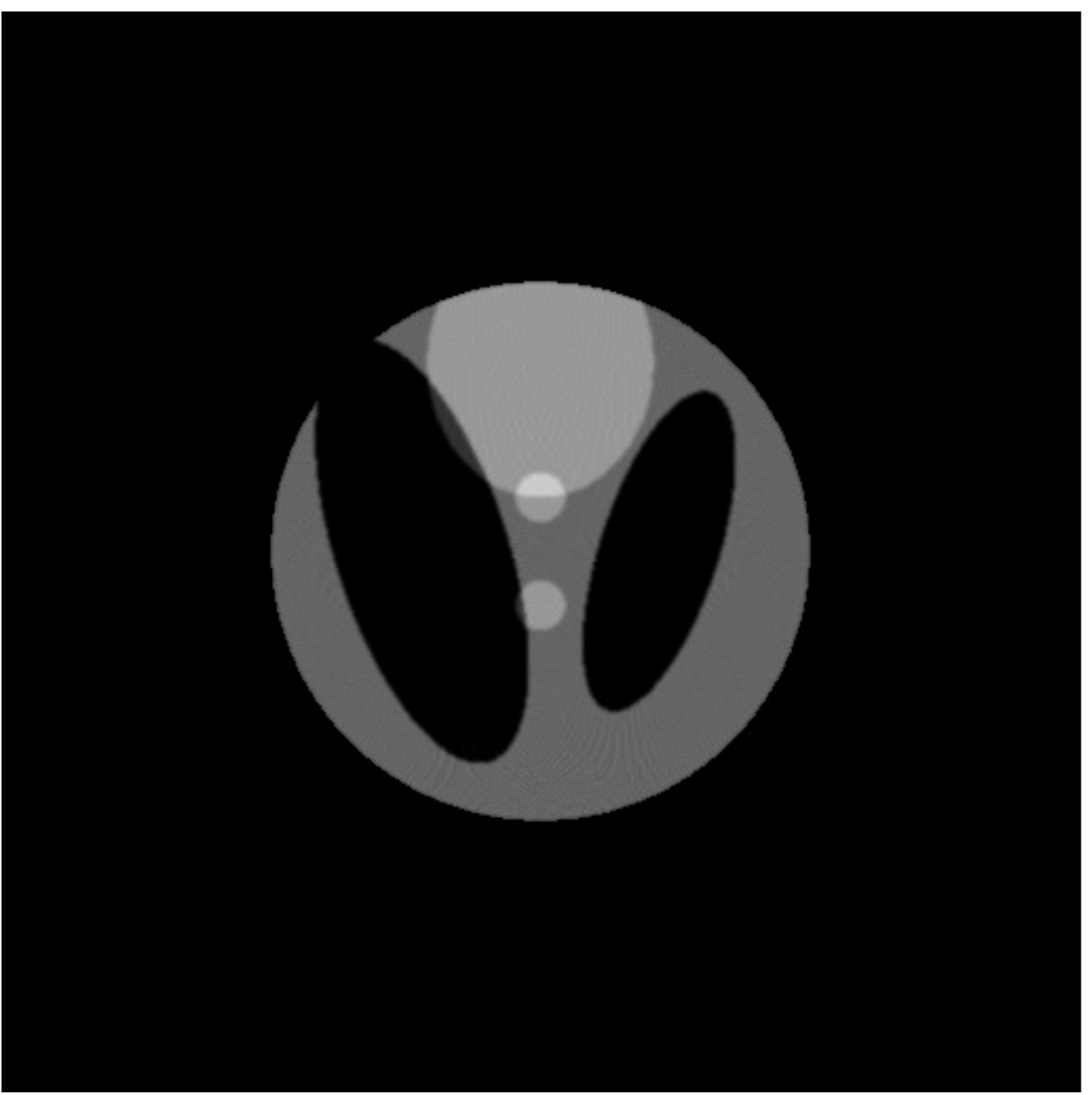}\label{fig:fig4_d}

} \subfigure[]{

\includegraphics[width=0.2\textheight]{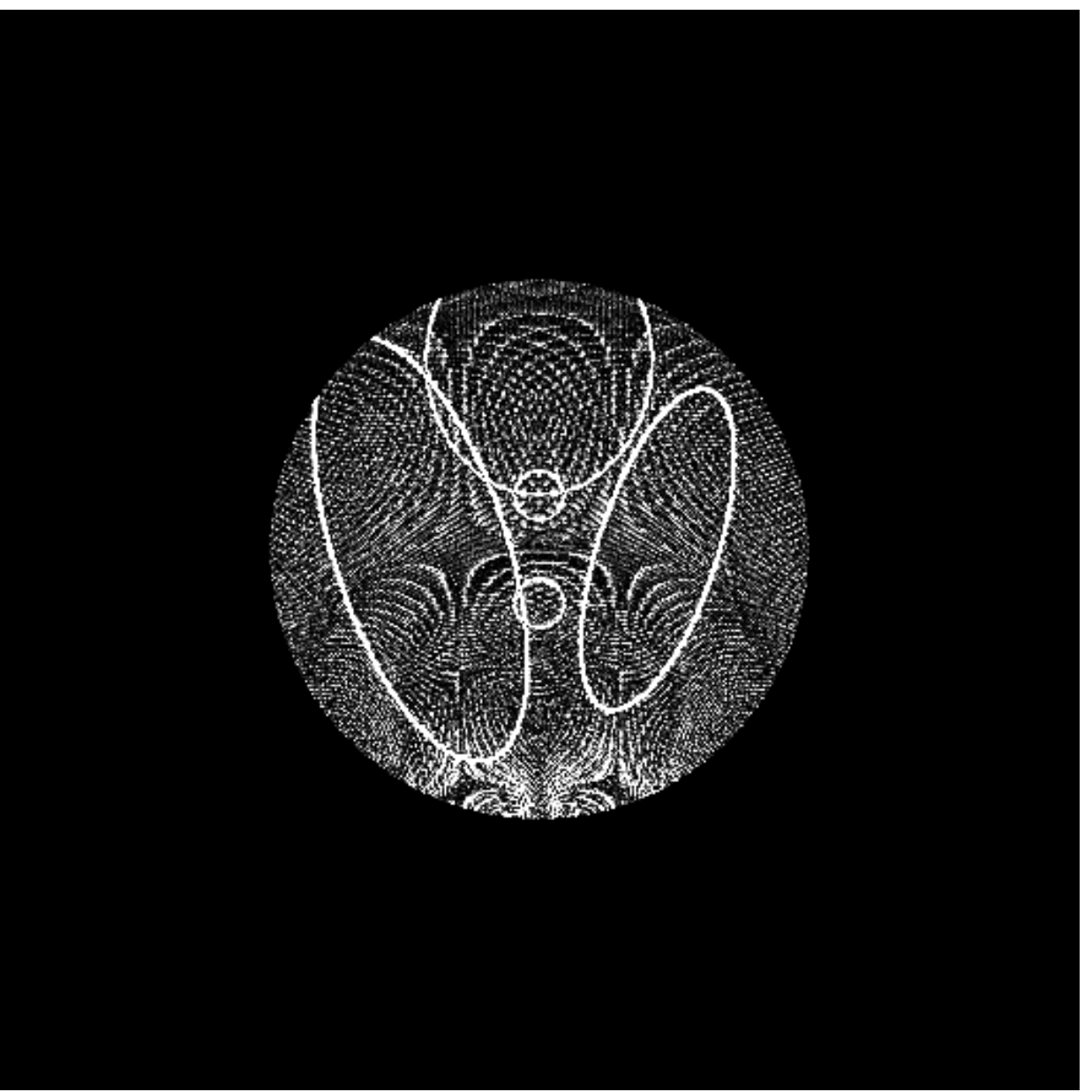}\label{fig:fig4_e}

} \subfigure[]{

\includegraphics[width=0.2\textheight]{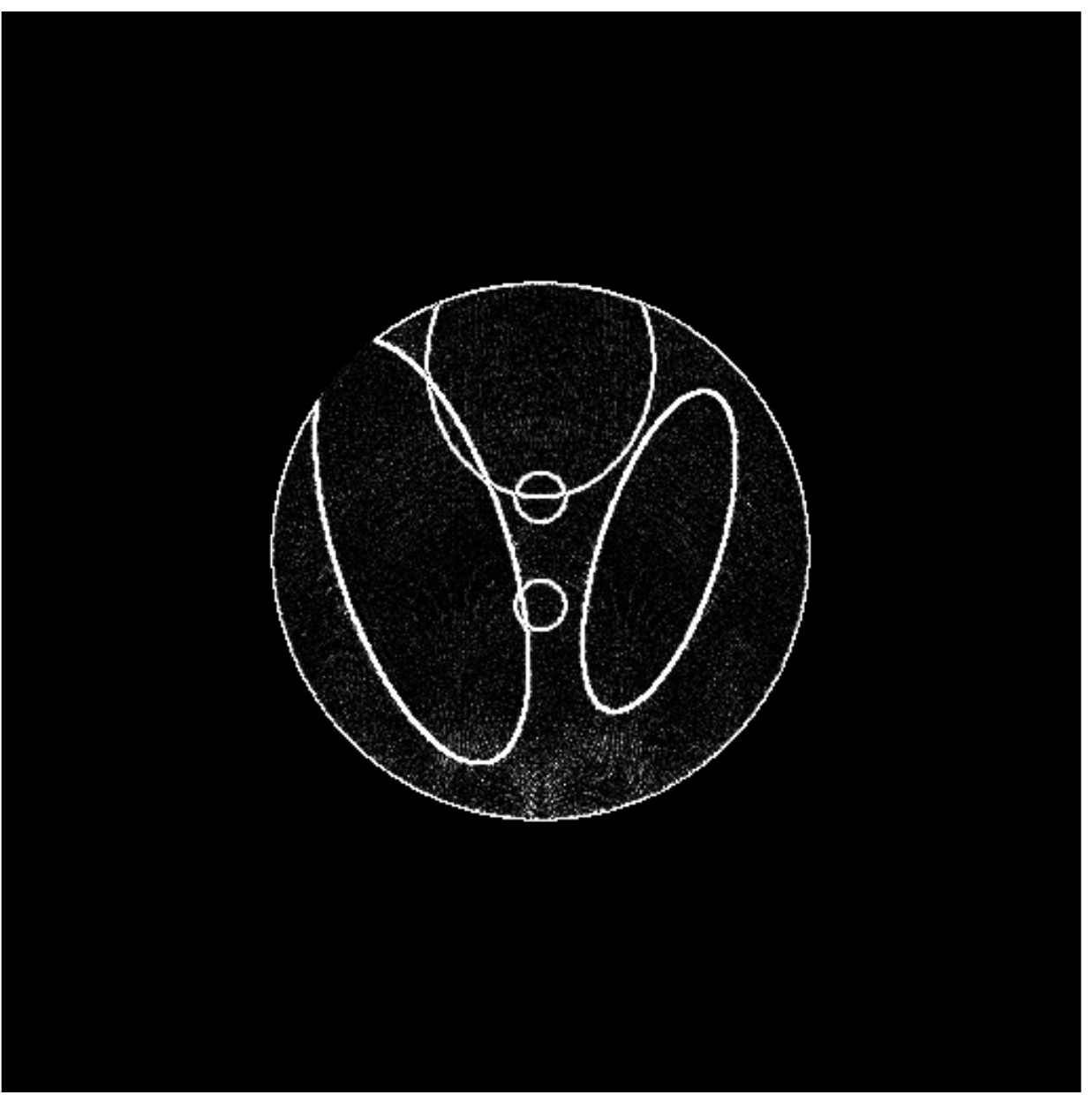}\label{fig:fig4_f}

}

\caption{(a) is image of the modified Shepp-Logan head phantom. (b) is crop
of the image of phantom corresponding to the ROI. (c) is the reconstruction
with FBP algorithm. (d) is the reconstruction with the iterative algorithm
of Eq.(\ref{eq:5-7}). (e) is the image of the errors for the FBP
method or NIRM. (f) is the image of errors for the iterative algorithm
of Eq. (\ref{eq:5-7}). }

\label{fig:fig4}
\end{figure}

In the following example the modified Shepp-Logan head phantom is
taken in consideration. A massive object is added outside the region
of interest. This massive object represents the bone of a human arm.
This object can introduce more artifacts for the reconstruction process.
The improvement of the iterative reconstruction can be seen more clearly
from this example. The stripe artifacts of FBP algorithm are remarkably
reduced for this example; see Figure~\ref{fig:fig4}. 

\subsection{The method}

\begin{figure}
 \centering

\subfigure[]{

\includegraphics[width=0.31\textheight]{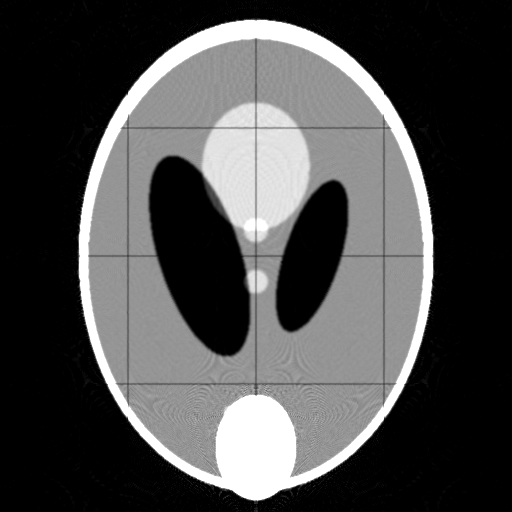}\label{Flo:fig25a} 

}\subfigure[]{

\includegraphics[width=0.27\textheight]{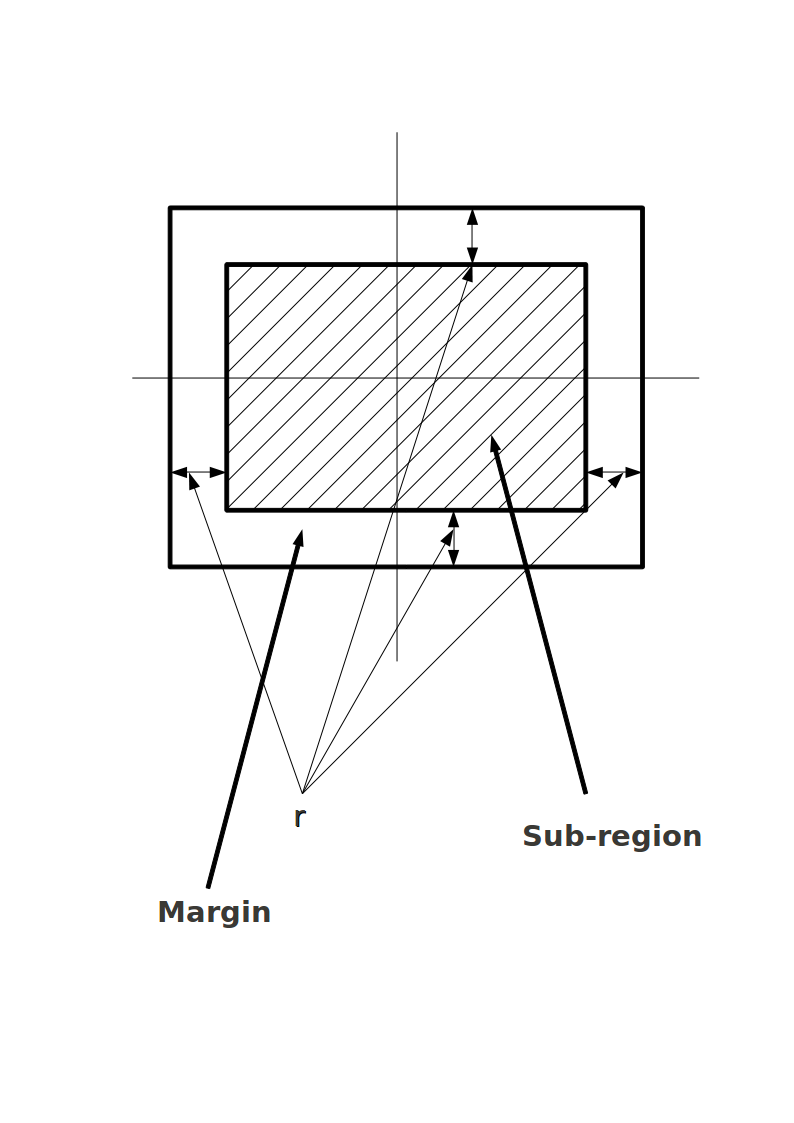}\label{Flo:fig25b} 

}

\caption{(a) the reconstructed image using Eq.(\ref{eq:20-3}). (b) Shows the
sub-region and the margin. The margin size is $r$.}

\label{fig:fig25}
\end{figure}

SIRM is sourced from the iterative reconstruction and re-projection
algorithm or local inverse method\cite{Ref-22-Shuangren-Zhao}, \cite{Ref-23-ShuangrenZhao}
for LFOV. However it requires an improvement to reconstruct the whole
image. First, the region of interest (ROI) can be made in any shapes.
It is not required that ROI is a round disk-shape region. In the past
ROI was chosen as round disk-shape region, this is because of the
situation of LFOV. In the following, ROI will be chosen as many small
square. The iterative algorithm\cite{Ref-22-Shuangren-Zhao} is used
to every square. The extrapolation is taken away because of FFOV.
The algorithm was posted on-line in Chinese roughly, see reference
\cite{Ref-25-shuangRenZhao} and it is summarized more details in
the following,
\begin{eqnarray}
p_{i}^{(1)} & = & p-P\,H_{i}\,X^{(0)}\label{eq:20-2}\\
X^{(1)} & = & \sum_{i=1}^{M}T_{i}\,R\,p_{i}^{(1)}\label{eq:20-3}
\end{eqnarray}
where $X^{(0)}$ is obtained in Eq.(\ref{eq:0-20}); where the subscript
$i$ is the index of sub-region which is a small square box; $M$
is the number of sub-regions. $X^{(1)}$ is the iterative reconstruction.
Superscript $(1)$ is corresponding to first iteration (the second
reconstruction). $p_{i}^{(1)}$ is iterative re-projection for $\omega_{i}$.
$\omega_{i}$ is the $i^{th}$ sub-region. After the parts of the
object in all sub-regions are reconstructed, all the parts of image
are put together to form the reconstructed image $X^{(1)}$. Two truncation
operators above are defined in the following
\begin{eqnarray}
H_{i}(x) & = & \left\{ \begin{array}{cc}
0 & \textrm{if}\:x\in\omega_{i}\\
1 & \textrm{if}\:x\notin\omega_{i}
\end{array}\right.\label{eq:30}
\end{eqnarray}
\begin{eqnarray}
T_{i}(x) & = & \left\{ \begin{array}{cc}
1 & \textrm{if}\:x\in\omega_{i}\\
0 & \textrm{if}\:x\notin\omega_{i}
\end{array}\right.\label{eq:40}
\end{eqnarray}
It is worthwhile to say that the operation $T_{i}(x)$ and $H_{i}(x)$
are normal multiplication. They are different from the operation of
$J$ which is corresponding to matrix multiplication. $T_{i}(x)$
is defined $1$ on a sub-region. $H_{i}(x)$ is defined $1$ in the
whole region except the sub-region. Hence $H_{i}(x)$ is a hole-shape
function. Define unitary operator

\begin{eqnarray}
1(x) & \equiv & 1\label{eq:50}
\end{eqnarray}
Thus, there is

\begin{eqnarray}
1(x) & =H_{i}(x) & +T_{i}(x)\label{eq:60}
\end{eqnarray}
Using above equation, Eq.(\ref{eq:20-2}) can be written as

\begin{eqnarray}
p_{i}^{(1)} & = & P\,T_{i}\,X^{(0)}+p-P\,X^{(0)}\label{eq:70}
\end{eqnarray}
However the results of Eq.(\ref{eq:20-2},\ref{eq:20-3}) shows cracks
between sub-regions. The cracks can be seen in the reconstructed image,
see Figure \ref{Flo:fig25a}. In order to eliminate the cracks, the
Eq.(\ref{eq:20-2} or \ref{eq:70}) is upgraded as

\begin{eqnarray}
p_{i}^{(1)} & = & P\,T_{i}^{+}\,X^{(0)}+p-P\,X^{(0)}\label{eq:80}
\end{eqnarray}
where

\begin{eqnarray}
T_{i}^{+}(x) & = & \left\{ \begin{array}{cc}
1 & \textrm{if}\:x\in\omega_{i}+\mathrm{Margin})\\
0 & \textrm{if}\:x\notin\omega_{i}+\mathrm{Margin})
\end{array}\right.\label{eq:90}
\end{eqnarray}
$T_{i}^{+}$ is the image truncation operator with margin, see Figure
\ref{Flo:fig25b}. It was found that if margin size $r$ is taken
as $4\sim10$ pixels, the cracks can be eliminated. But the margin
can be chosen as for example 40 if the image size is big. Considering

\begin{equation}
\sum_{i=1}^{M}T_{i}\,f=f\label{eq:91}
\end{equation}
and substituting Eq.(\ref{eq:80}) to Eq.(\ref{eq:20-3}), there is

\begin{equation}
X^{(1)}=(\sum_{i=1}^{M}T_{i}\,R\,P\,T_{i}^{+}X^{(0)})+R\,(p-P\,X^{(0)})\label{eq:100}
\end{equation}
or

\begin{equation}
X^{(1)}=[(\sum_{i=1}^{M}T_{i}\,R\,P\,T_{i}^{+})+(I-J)]X^{(0)}\label{eq:110}
\end{equation}
where $J=R\:P$ defined in Eq.(\ref{eq:0-40}). The above formula
can be rewritten as,

\begin{equation}
X^{(1)}=F_{SIRM}\,X^{(0)}\label{eq:120}
\end{equation}
$F_{SIRM}$ is a filtering function corresponding to SIRM. The filter
function is defined as

\begin{equation}
F_{SIRM}\equiv U+(I-J)\label{eq:130}
\end{equation}
where $U$ is the sub-region projection and reconstruction operator 

\begin{equation}
U\equiv\sum_{i=1}^{M}T_{i}\,J\,T_{i}^{+}\label{eq:140}
\end{equation}
Corresponding to Eq.(\ref{eq:130}), there is, 

\begin{equation}
f_{SIRM}(y,x)=u(y,x)+\delta(y,x)-j(y,x)\label{eq:150-a}
\end{equation}
The above filtering function does not satisfy the unitary condition
which keeps the dc value unchanged after the reconstruction compare
to the original image. Hence it is required to be upgraded as

\begin{equation}
f_{SIRM}(y,x)=u(y,x)+\eta(x)\,\delta(y,x)-j(y,x)\label{eq:150}
\end{equation}
$\eta(x)$ is normalization function similar to which used in LIRM.
Considering the unitary condition

\begin{equation}
\sum_{y}f_{SIRM}(y,x)=1\label{eq:160}
\end{equation}
implies

\begin{equation}
\eta(x)=2-\sum_{y}u(y,x)\label{eq:170}
\end{equation}
Here the summation is taken on the definition area of the variable
$y\,\in V$. $V=\sum_{i}\omega_{i}$ is the region of whole image.
The Eq.(\ref{eq:130}) can be replaced as

\begin{equation}
F_{SIRM}\equiv\eta I-J+U\label{eq:130A}
\end{equation}
Usually $\eta(x)$ can be taken as

\begin{equation}
\eta(x)\approx1\label{eq:180}
\end{equation}
Only if the sub-region ($\omega_{i}$) is very small, $\eta(x)$ is
possible to be significantly different from $1$. This is same as
the case of the LIRM discussed in last section. The resolution function
of SIRM algorithm is $F_{SIRM}\:J$. 

\begin{figure}
 \centering

\subfigure[]{

\includegraphics[width=0.17\textheight]{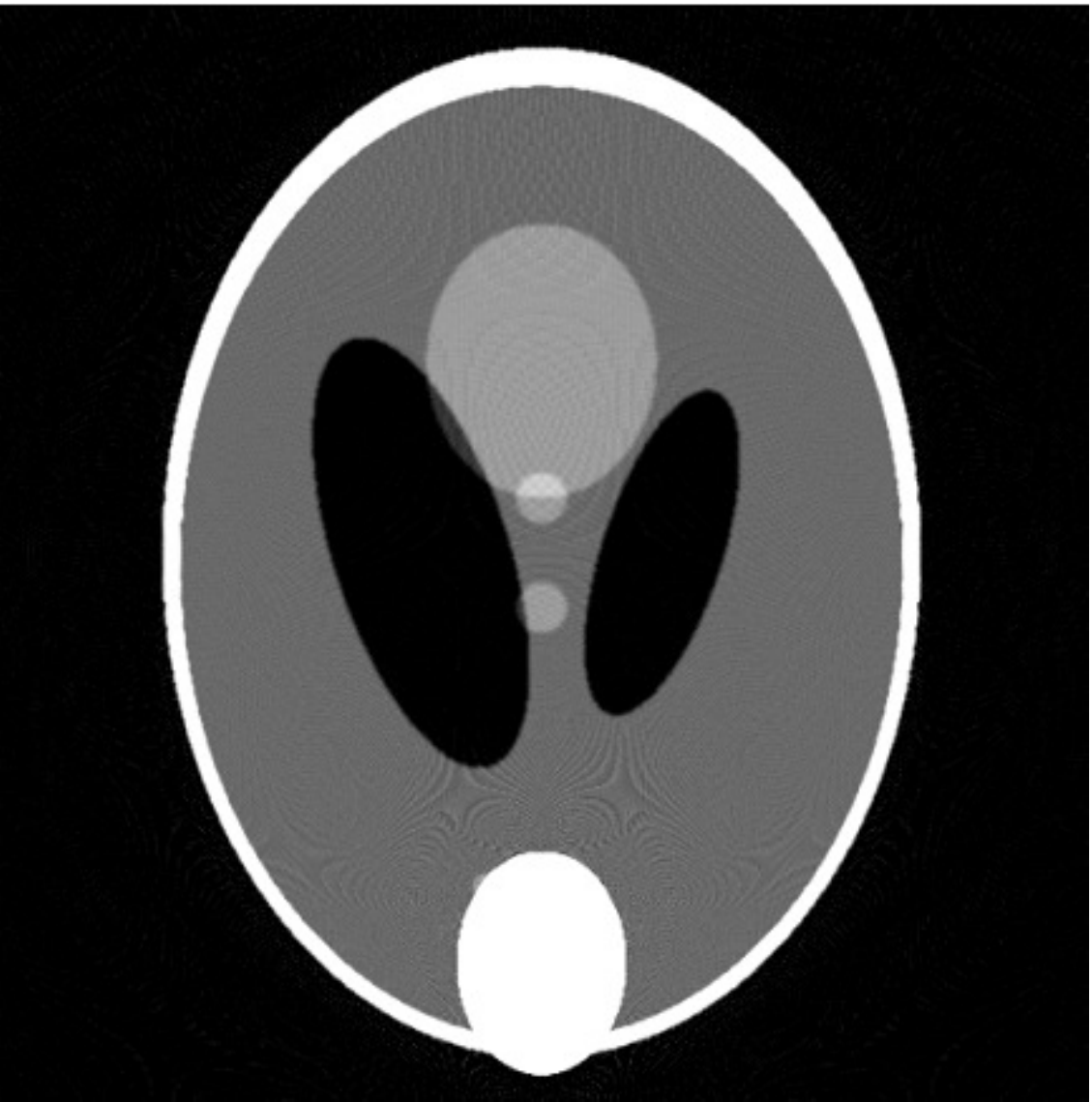}\label{fig:fig8_a}

} \subfigure[]{

\includegraphics[width=0.15\textheight]{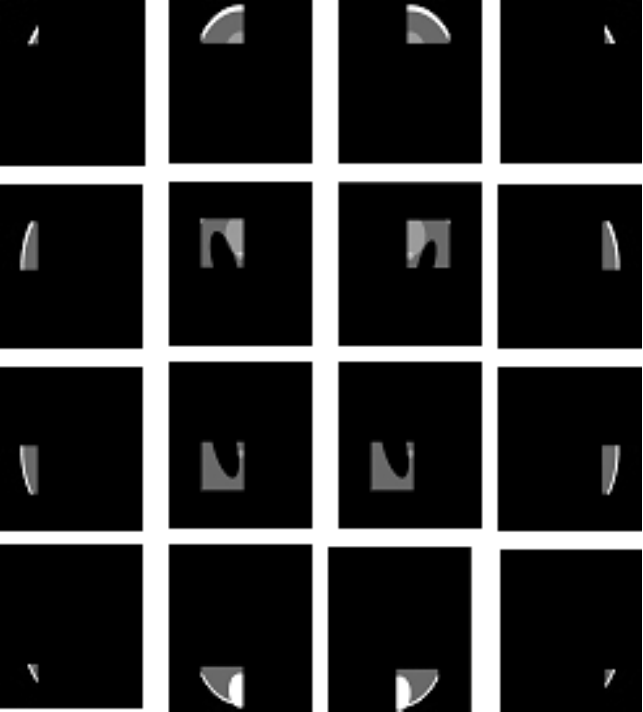}\label{fig:fig8_b}

} \subfigure[]{

\includegraphics[width=0.13\textheight]{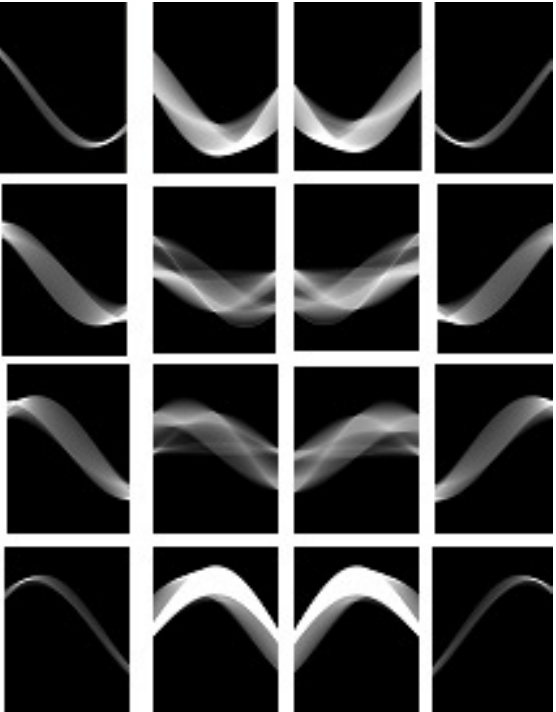}\label{fig:fig8_c}

} 

\subfigure[]{

\includegraphics[width=0.15\textheight]{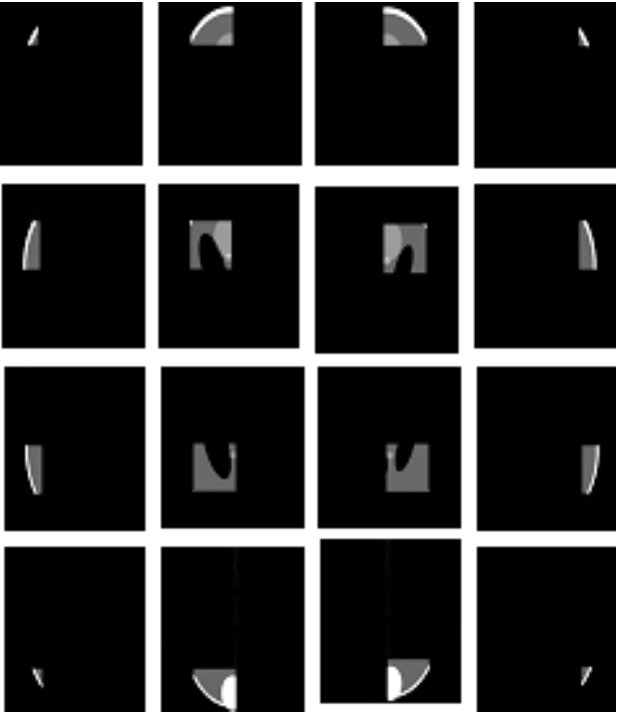}\label{fig:fig8_d}

} \subfigure[]{

\includegraphics[width=0.17\textheight]{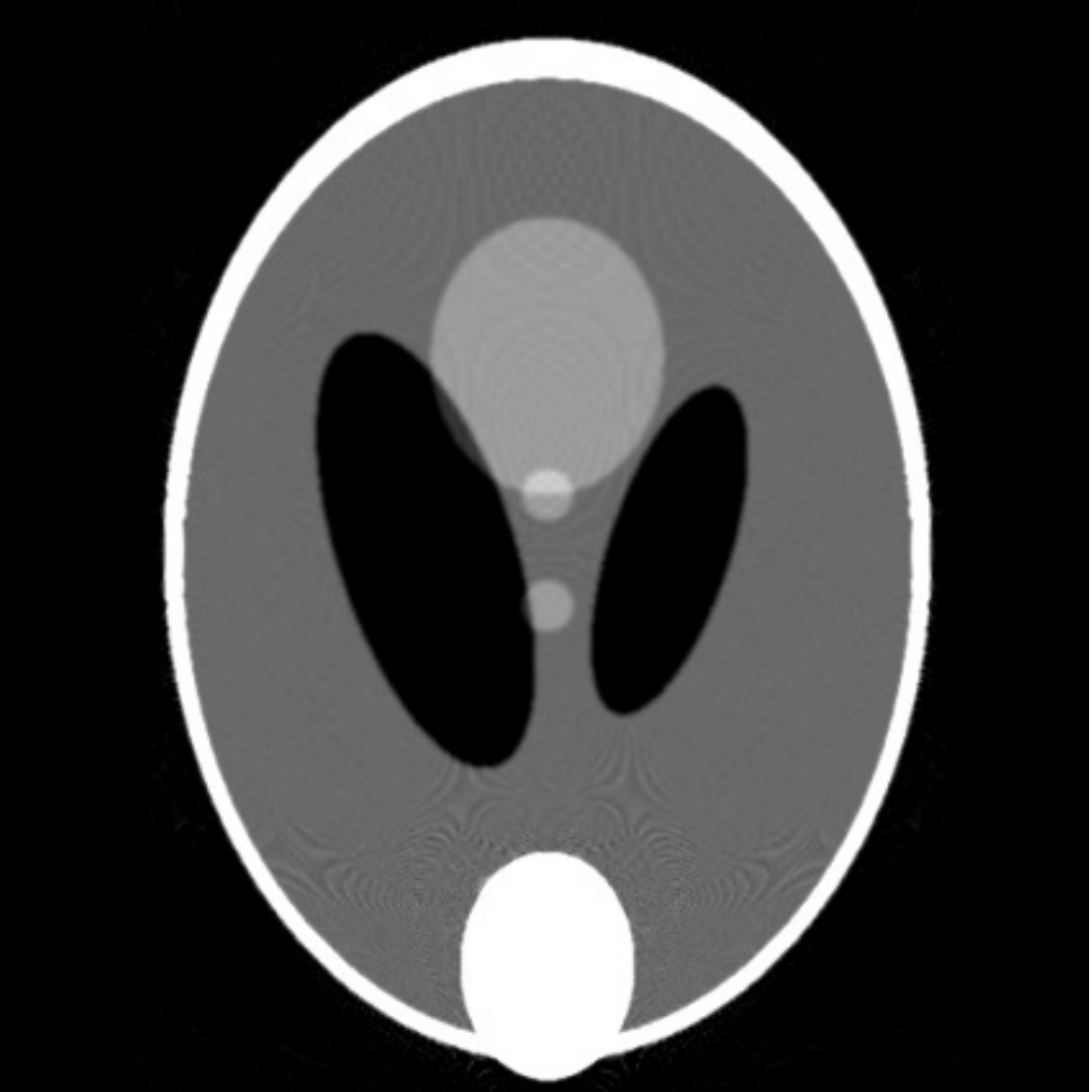}\label{fig:fig8_e}

} 

\caption{$U\,X^{(0)}=\sum_{i=1}^{M}T_{i}\,R\,P\,T_{i}^{+}X^{(0)}$. (a) The
first reconstruction. (b) The image of (a) is divided in sub-regions.
(c)The sub-region images of (b) are reprojected. (d) The sub-region
images are reconstructed from (c). (e) The sub-region images are put
together to form a whole image: $U\,X^{(0)}$. }

\label{fig:fig8}
\end{figure}

The biggest difference of the form of SIRM from TIRM is the operator
$U$ in Eq.(\ref{eq:140}) comparing Eq.(\ref{eq:130A}) and Eq.(\ref{eq:4}).
In order to have a good understanding of this operator, the process
of this operator is shown in Fig.~\ref{fig:fig8}. Fig.~\ref{fig:fig8_a}
illustrates the first reconstruction $X^{(0)}=R\,p$. Fig.~\ref{fig:fig8_b}
shows that the image of Fig.~\ref{fig:fig8_a} is divided into sub-regions
$T_{i}^{+}\,X^{(0)}\;i=1,2,....M$. Fig.~\ref{fig:fig8_c} shows
that the sub-region images of (b) are reprojected $P\,T_{i}^{+}\,X^{(0)}\;i=1,2,....M$.
Fig.~\ref{fig:fig8_d} shows that the sub-region image is reconstructed
from Fig.~\ref{fig:fig8_c} by using $R\,P\,T_{i}\,X^{(0)}\;i=1,2,....M$.
Fig.~\ref{fig:fig8_e} shows that the sub-region images are put together
to form a whole image by using $U\,X^{(0)}=\sum_{i=1}^{M}T_{i}\,R\,P\,T_{i}^{+}X^{(0)}$. 

The original image Fig.~\ref{fig:fig8_a} is chosen as the modified
Shepp-Logan head phantom with data size $512\times512$. The modification
is adding a massive small disk to the bottom of the image. The massive
small disk will increase the normal artifacts, which will be utilized
to test the algorithms in the next paragraph. The projection operator
$P$ is parallel beam and defined in Matlab. The projections data
is created by the operator $P$ to the above modified Shepp-Logan
head phantom. The number of projections is 360 for the half circle
scan (180 degree); the space between the two elements of the detector
is taken equal to the space between the two pixels of image. The operator
$R$ is corresponding to NIRM which is filtered back projection method
which is defined in Matlab. 

The process of operator $U$ shown in Fig.~\ref{fig:fig8} looks
mediocre. Actually, it is really not mediocre because the margins
in the algorithm plays an important role in eliminating artifacts
and decreasing the noises.

In the limit case $\omega_{i}$ is small as only one pixel, 

\begin{equation}
\sum_{i=1}^{M}T_{i}=1(x)\label{eq:181}
\end{equation}

\begin{equation}
J\,T_{i}^{+}=K\label{eq:182}
\end{equation}
Here $K$ is defined in Eq.(\ref{eq:sec4-01}), Hence considering
Eq.(\ref{eq:140}), there is 

\begin{equation}
U=K\label{eq:183}
\end{equation}
Considering Eq.(\ref{eq:130A}) Eq.(\ref{eq:sec40-100}) there is
 SIRM$\rightarrow$LIRM in the case $\omega_{i}\rightarrow$one pixel.

\subsection{The iterative algorithm with more loops}

In the above discussion, two algorithms are iterated with only one
loop. If one loop does not satisfy, more loops can be utilized, this
can be written as

\begin{equation}
X_{l}^{(n)}=F_{l}^{n}\,X^{(0)}\label{eq:190}
\end{equation}
where $X_{l}^{(n)}$ is the reconstructed image with more loops of
iteration. $n$ is the iteration number. $F_{l}^{n}=(F_{l})^{n}$
is the filtering operator for iteration number $n$. $l$ indicates
different algorithm, $l=\{TIRM,\,LIRM,\,SIRM\}$. The resolution function
with more loops is

\begin{equation}
J_{l}^{(n)}=F_{l}^{n}J\label{eq:195}
\end{equation}
For the reconstructions, the error is defined as

\begin{equation}
Err_{l}^{(n)}\equiv X_{l}^{(n)}-X_{o}\label{eq:200}
\end{equation}
 $X_{l}^{(n)}$ is defined in Eq.(\ref{eq:190}). $X_{o}$ is the
object or the original image. Considering Eq.(\ref{eq:0-20}) and
Eq.(\ref{eq:190}) the error can be defined as

\begin{equation}
Err_{l}^{(n)}=(F_{l}^{n}J-I)X_{o}+F_{l}^{n}\,R\,p_{n}\label{eq:205}
\end{equation}
The first item of the above formula is corresponding to artifacts
which is related to the original image $X_{o}$; the second item is
corresponding to noises which is related to noises in the first reconstruction
$R\,p_{n}$. The error for NIRM method is $Err^{(0)}=(J-I)X_{o}-R\:p_{n}$.
The artifact transfer function can be defined as 
\begin{equation}
A_{l}^{n}=F_{l}^{n}J-I\label{eq:206}
\end{equation}
and the noise transfer function can be defined as 
\begin{equation}
N_{l}^{n}=F_{l}^{n}\:R\label{eq:207}
\end{equation}
The noise transfer function and the artifact transfer function have
different forms, which gives the possibility to optimize the algorithm
by balancing the artifacts and the noises and adjusting the filtering
function. 

The absolute error $|Err_{l}^{(n)}|$ will be used to study reconstruction
results. The distance between the reconstructed image $X_{l}^{(n)}$
and the original image $X$ can be used also to compare the reconstruction.
The distance is defined in the following, 

\begin{equation}
d_{l}^{(n)}=\frac{\sum_{x}(X_{l}^{(n)}(x)-X_{o}(x))^{2}}{\sum_{x}(X(x)-\bar{X}_{o}(x))^{2}}\label{eq:210}
\end{equation}
where $\bar{X}$ is the average of the image $X(x)$. See reference\cite{Ref-5-Paul-S-Cho}
for details of the definition of the distance.

\subsection{Results}

\begin{figure}
 \centering

\subfigure[]{

\includegraphics[width=0.2\textheight]{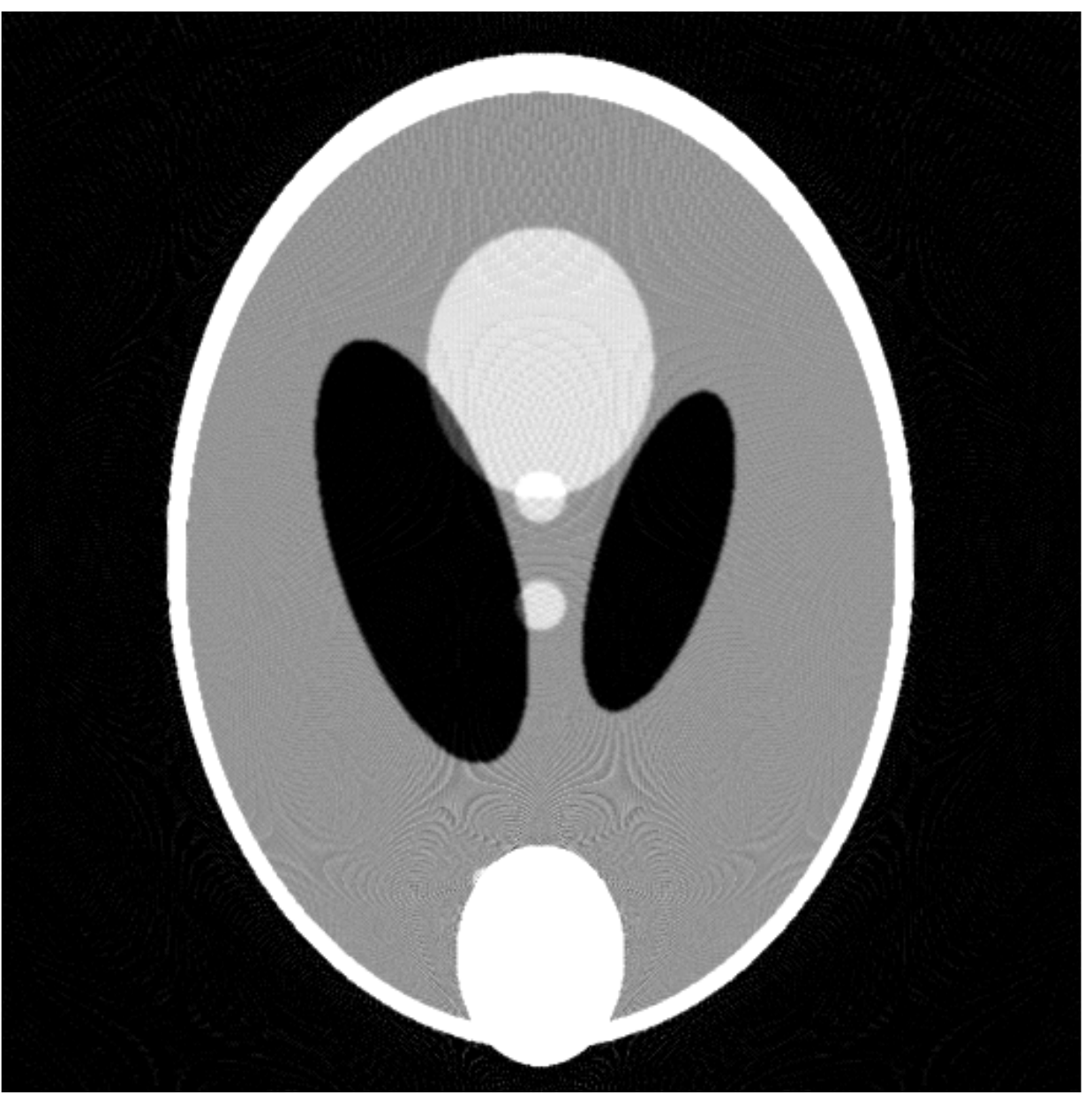}\label{fig:fig6_a}

}\subfigure[]{

\includegraphics[width=0.2\textheight]{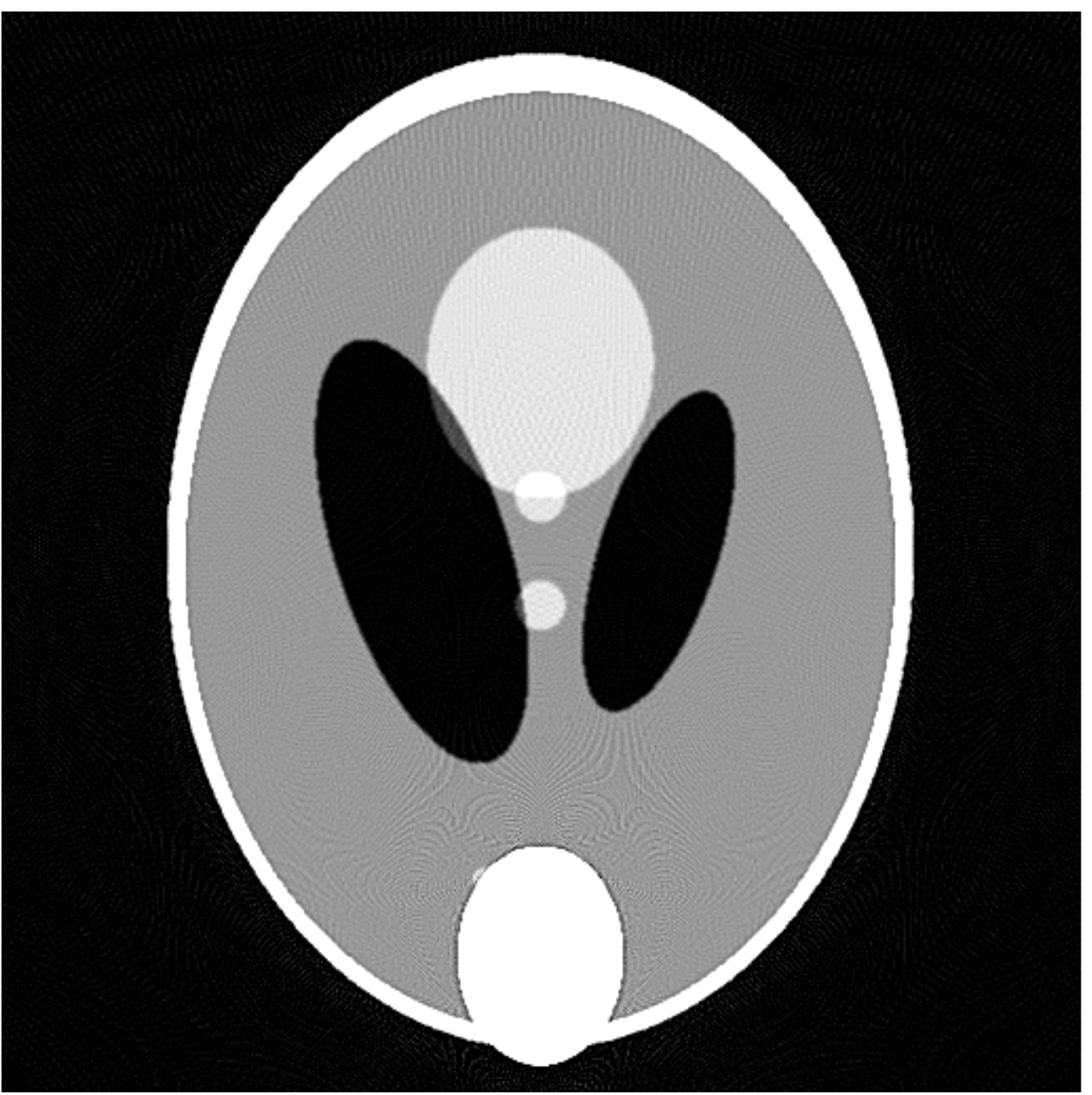}\label{fig:fig6_b}

}\subfigure[]{

\includegraphics[width=0.2\textheight]{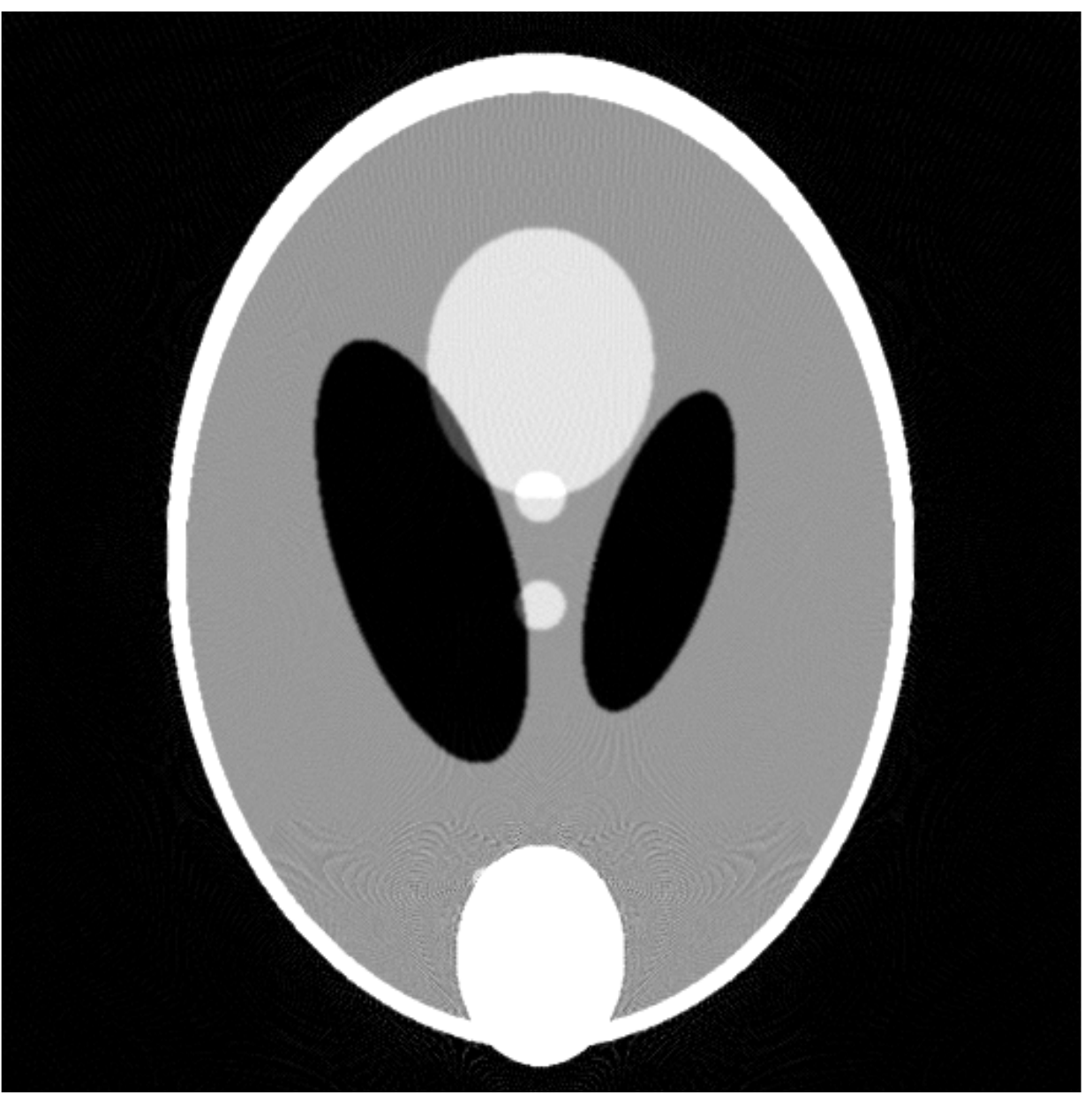}\label{fig:fig6_c}

}

\subfigure[]{

\includegraphics[width=0.2\textheight]{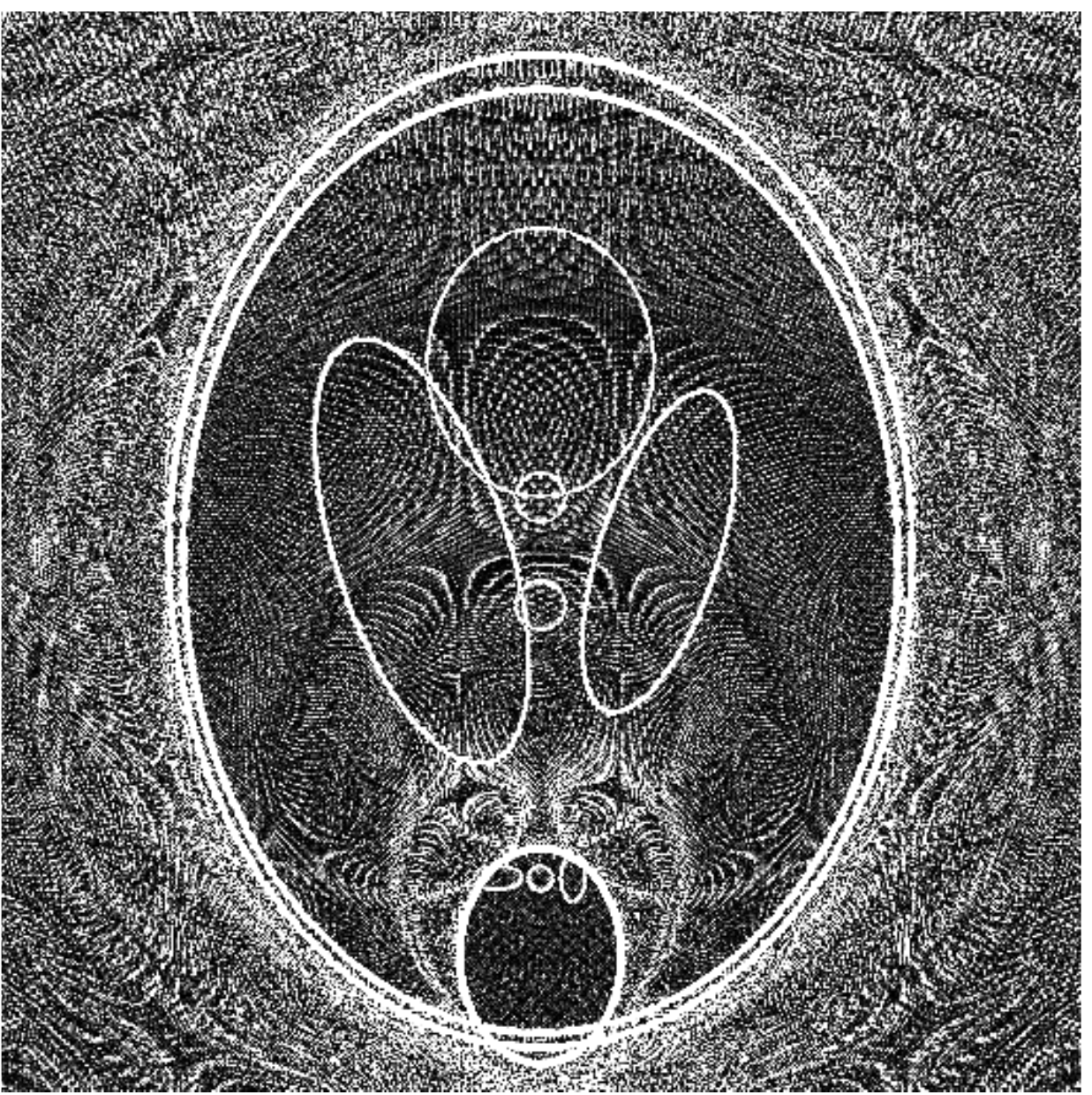}\label{fig:fig6_d}

}\subfigure[]{

\includegraphics[width=0.2\textheight]{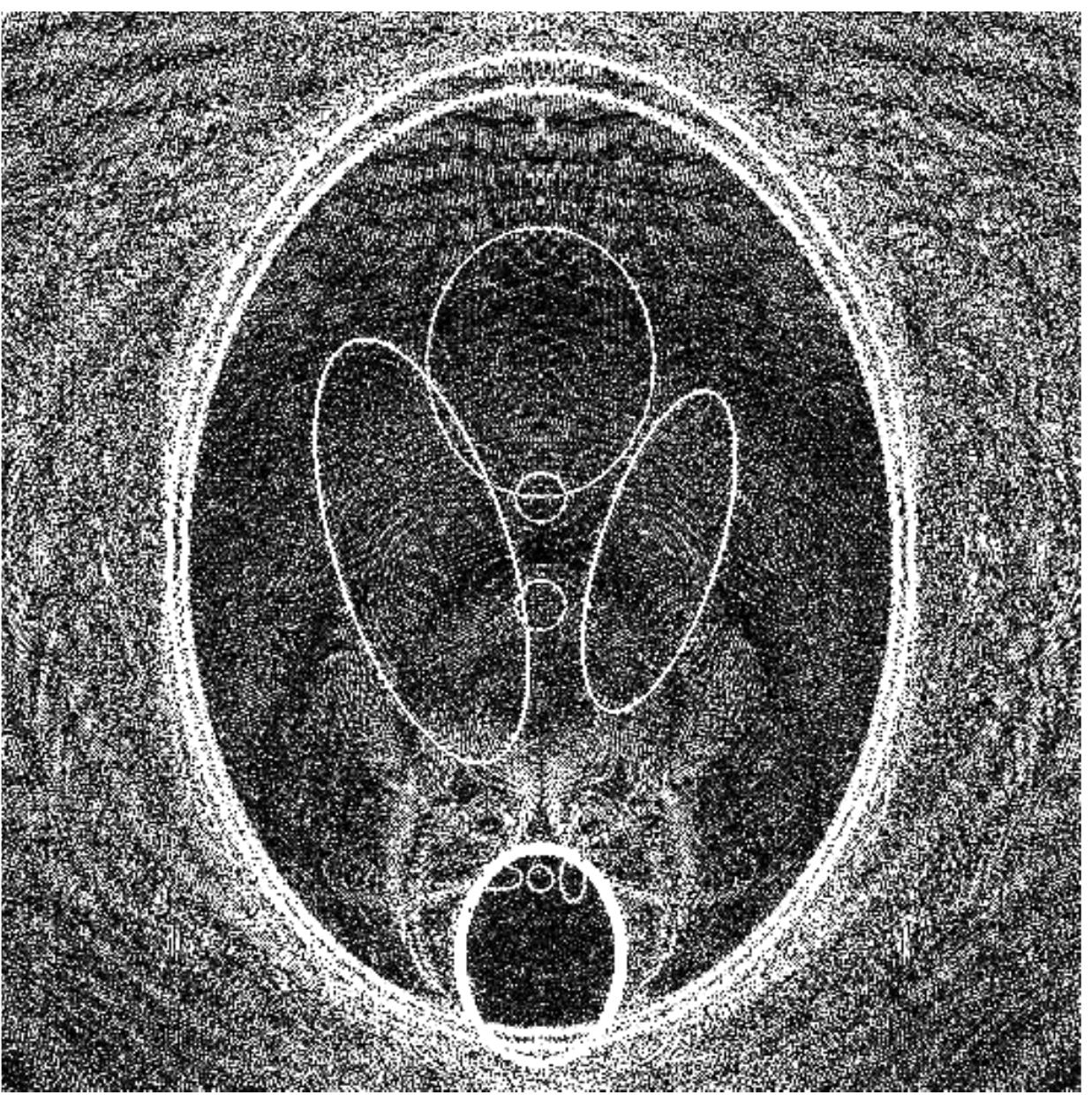}\label{fig:fig6_e}

}\subfigure[]{

\includegraphics[width=0.2\textheight]{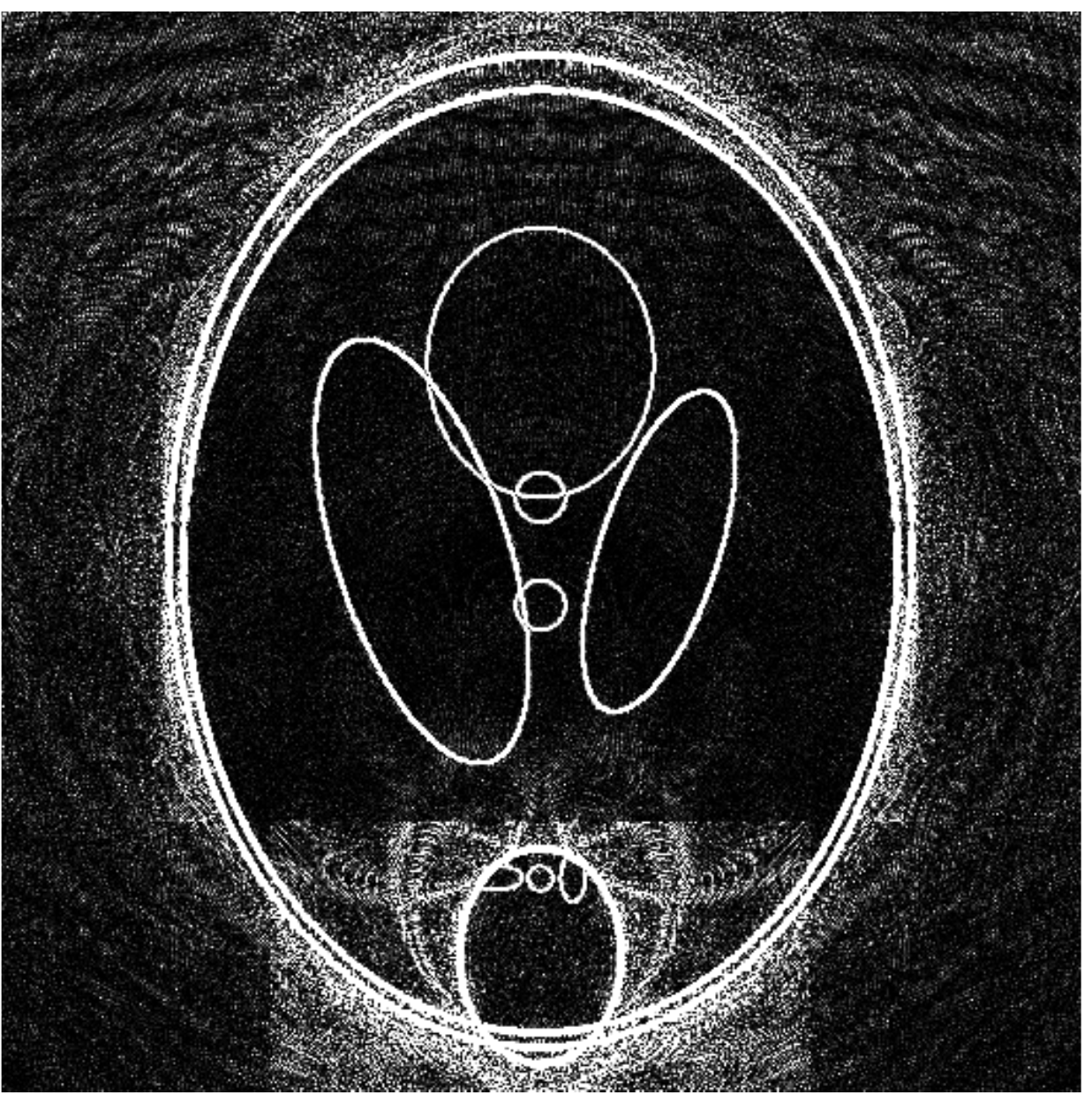}\label{fig:fig6_f}

} 

\subfigure[]{

\includegraphics[width=0.2\textheight]{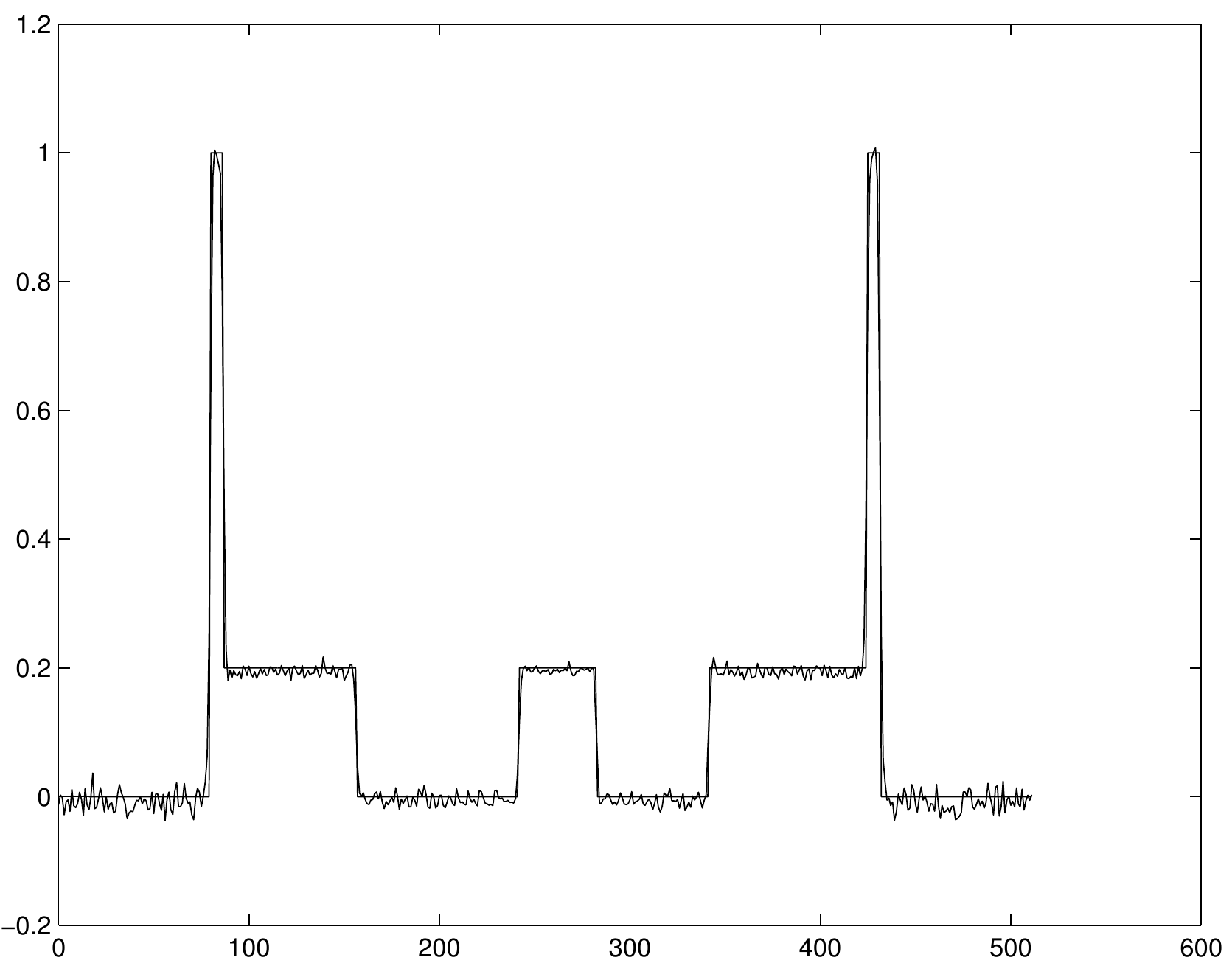}\label{fig:fig6_g}

}\subfigure[]{

\includegraphics[width=0.2\textheight]{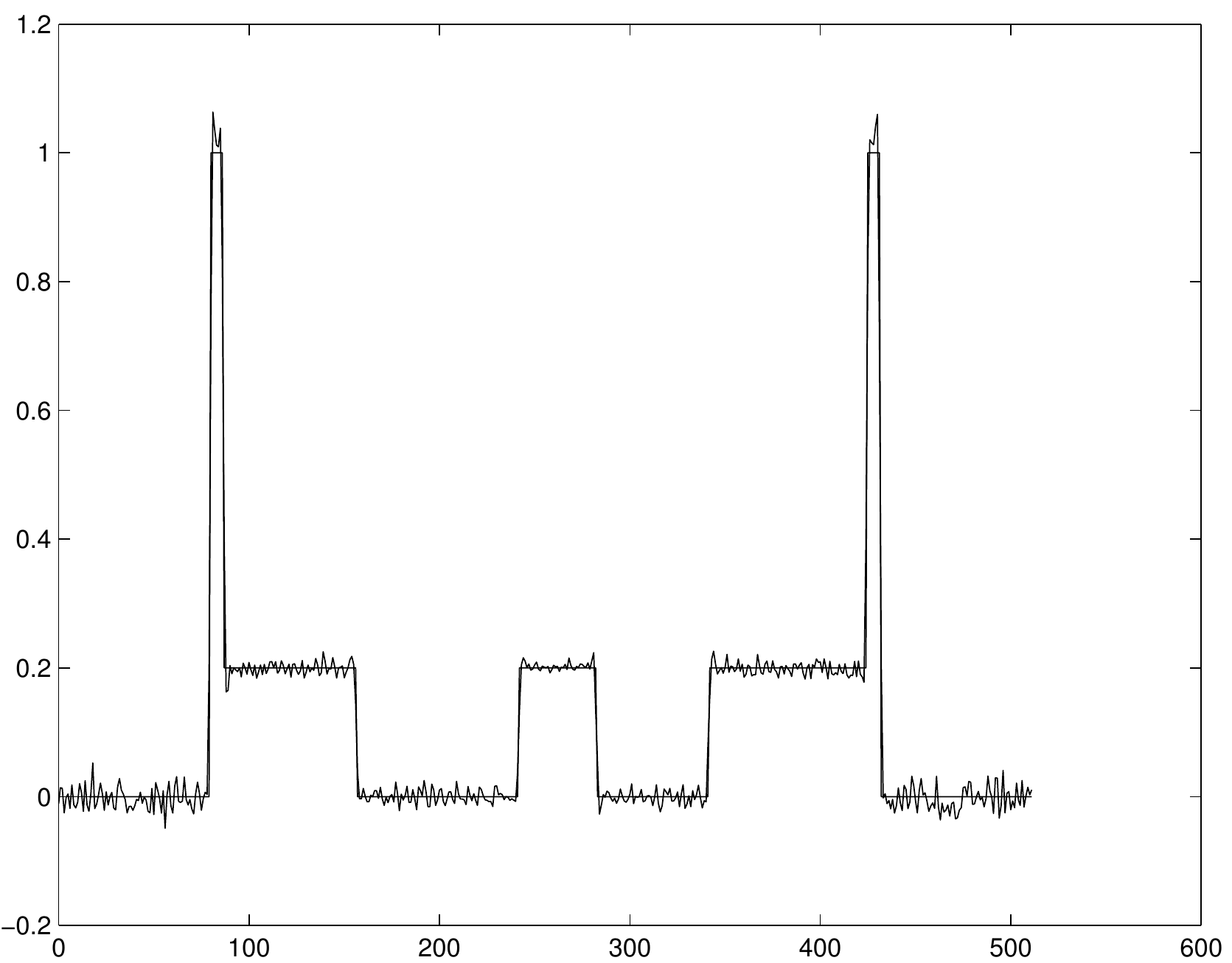}\label{fig:fig6_h}

}\subfigure[]{

\includegraphics[width=0.2\textheight]{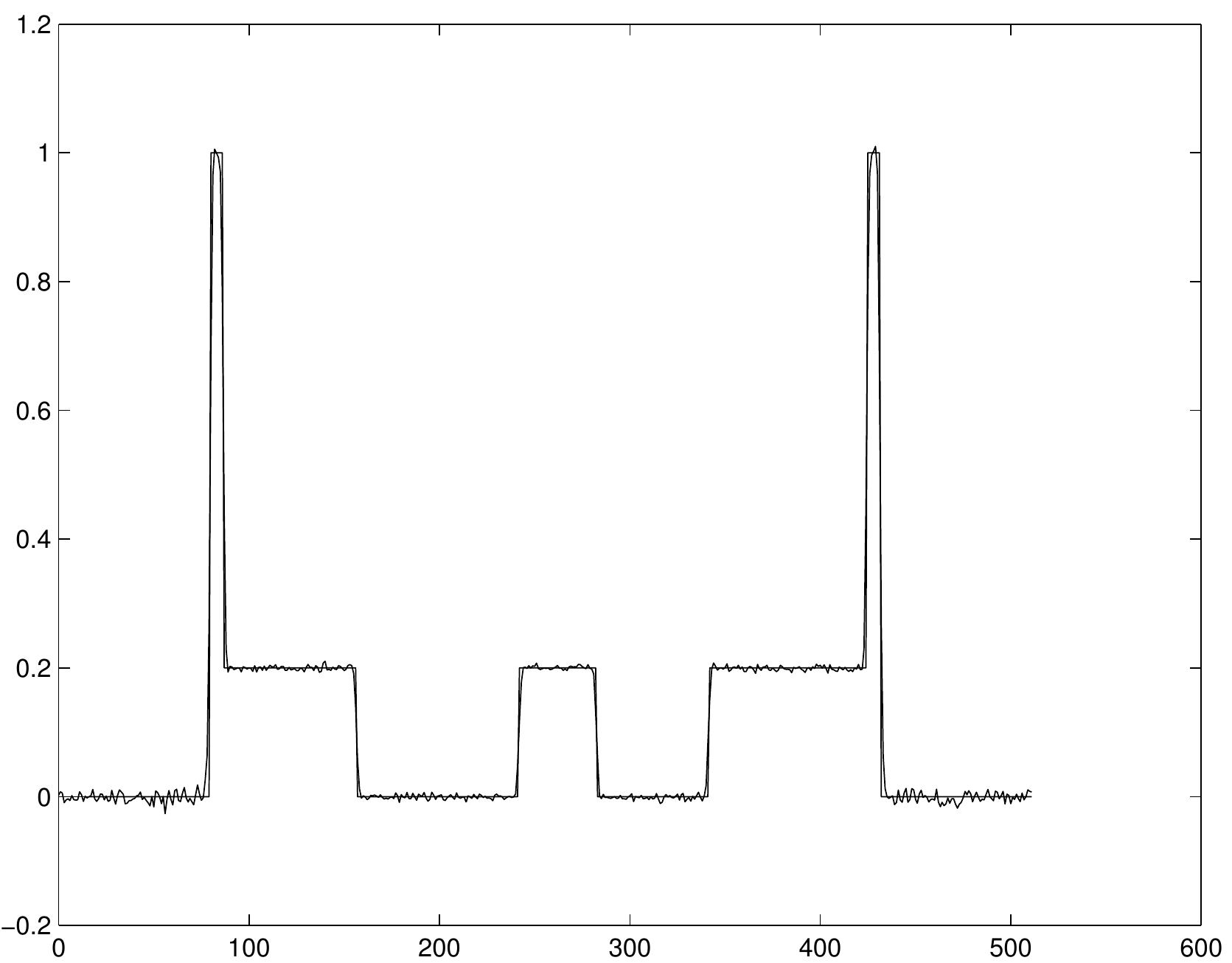}\label{fig:fig6_f-1}

}

\caption{The reconstructions of different algorithms. (a) The reconstruction
with NIRM. (b) The reconstruction with TIRM . (c) The reconstruction
with SIRM. (d), (e) and (f) are the absolute errors corresponding
to image (a), (b) and (c) respectively. (g), (h) and (i) are the profiles
along x-axis corresponding to image (a), (b) and (c).}

\label{fig:fig6}
\end{figure}

Figure~\ref{fig:fig6} shows the comparison of SIRM with NIRM and
TIRM. NIRM is implemented in Eq(). TIRM is implemented with Eq.(\ref{eq:2}).
$\eta$ in Eq.(\ref{eq:150}) is chosen as $1$ for simplification.
SIRM is implemented with Eq.(\ref{eq:110}). The region of the object
is divided according to $4\,\times\,4$ grid for SIRM. Hence, there
are $M=16$ sub-regions. The margin is chosen as $10$ pixels. The
iteration for Eq.(\ref{eq:190}) are done with only one loop, i.e.
$n=1$ . 

The original image $X_{o}$, the projection projector $P$ and the
reconstruction operator $R$ are chosen the same as in Fig.~\ref{fig:fig8},
see section 5.2. The projections $p$ is obtained through the simulation
with $p=P\,X$. In this example additional noises $p_{n}$ are not
added to the projections. However, since there are always calculation
errors, $p_{n}\neq0$ in general.

SIRM yielded the best reconstruction results than the NIRM and the
TIRM. The stripe artifacts shown on Figure~\ref{fig:fig6_a} are
reduced remarkably on Figure~\ref{fig:fig6_c}. The absolute errors
$|Err^{(0)}|$ shown in Figure~\ref{fig:fig6_d} $|Err_{l=NRRM}^{(0)}|$
and Figure~\ref{fig:fig6_e} $|Err_{l=TRRM}^{(1)}|$ are larger than
in Figure~\ref{fig:fig6_f} $|Err_{l=SRRM}^{(1)}|$. The results
of TIRM (Figure~\ref{fig:fig6_b},\ref{fig:fig6_e}) are similar
to the results of NIRM method (Figure~\ref{fig:fig6_a},\ref{fig:fig6_d}). 

It is important to mention that: A) in Figure~\ref{fig:fig6_f} the
absolute errors on the two sub-regions containing the massive disk
are little bit larger than the errors in other sub-regions. This drawback
can be eliminated through increasing the number of sub-regions, for
example using $16\times16$ grid instead of $4\times4$ grid. In practice,
the smaller sub-regions are required to be used only in the two sub-regions
containing the massive disk. B) the above results of the iterative
algorithm are only done with one loop of iteration and further more
loops using Eq.(\ref{eq:190}) can also improve the results, but the
improvement is limited. C) If LIRM is implemented, since the sub-region
becomes as small as only one pixel(voxel), the result should be better
(in the meaning of reducing the artifacts and decreasing the noises)
than SIRM if the same margin $r$ is used. LIRM is more time-consuming,
to implement it more modern technologies for example GPU parallel
calculation and fast back-projection techniques are required. The
implementation of LIRM will be left for the future work.

\begin{table}
\centering

\begin{tabular}{|c|c|}
\hline 
Methods: & Distance:\tabularnewline
\hline 
\hline 
NIRM Eq.(\ref{eq:0-20}) & 0.0177\tabularnewline
\hline 
TIRM Eq.(\ref{eq:2}) & 0.0134\tabularnewline
\hline 
SIRM Eq.(\ref{eq:110}) & 0.0172\tabularnewline
\hline 
\end{tabular}

\caption{The distance for different methods}

\label{tab:1}
\end{table}

The distances for the above three algorithms have been calculated.
Table \ref{tab:1} tells that the distance from the reconstruction
of SIRM to the phantom is smaller than the distance from the reconstruction
of NIRM to the phantom. However, the smallest distance is obtained
through TIRM. Do these results mean that the reconstruction results
from the TIRM is better than SIRM? The following details of the profiles
give the answer. 

\begin{figure}
 \centering

\subfigure[]{

\includegraphics[width=0.5\textheight]{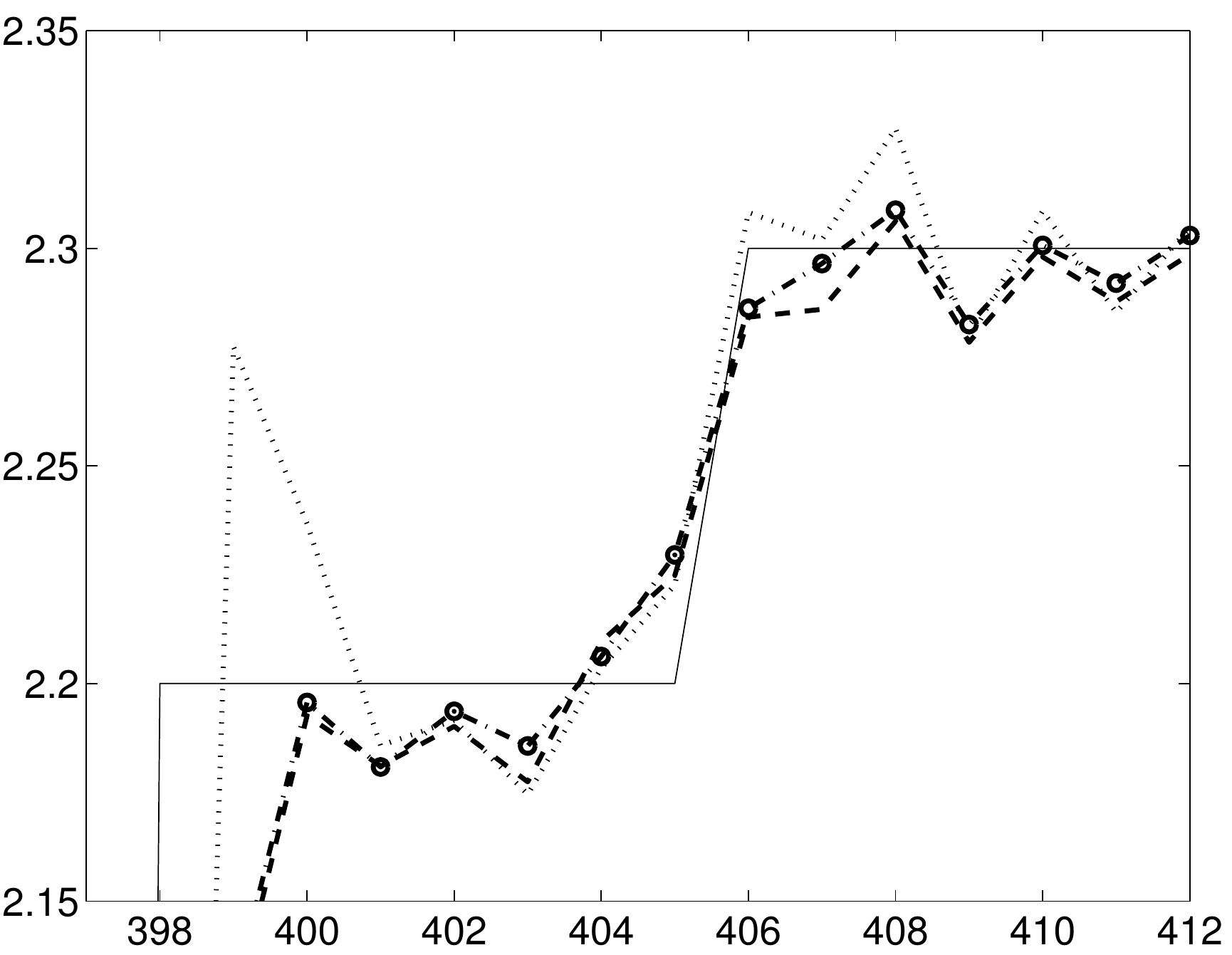}\label{fig:fig10_a}

}

\subfigure[]{

\includegraphics[width=0.5\textheight]{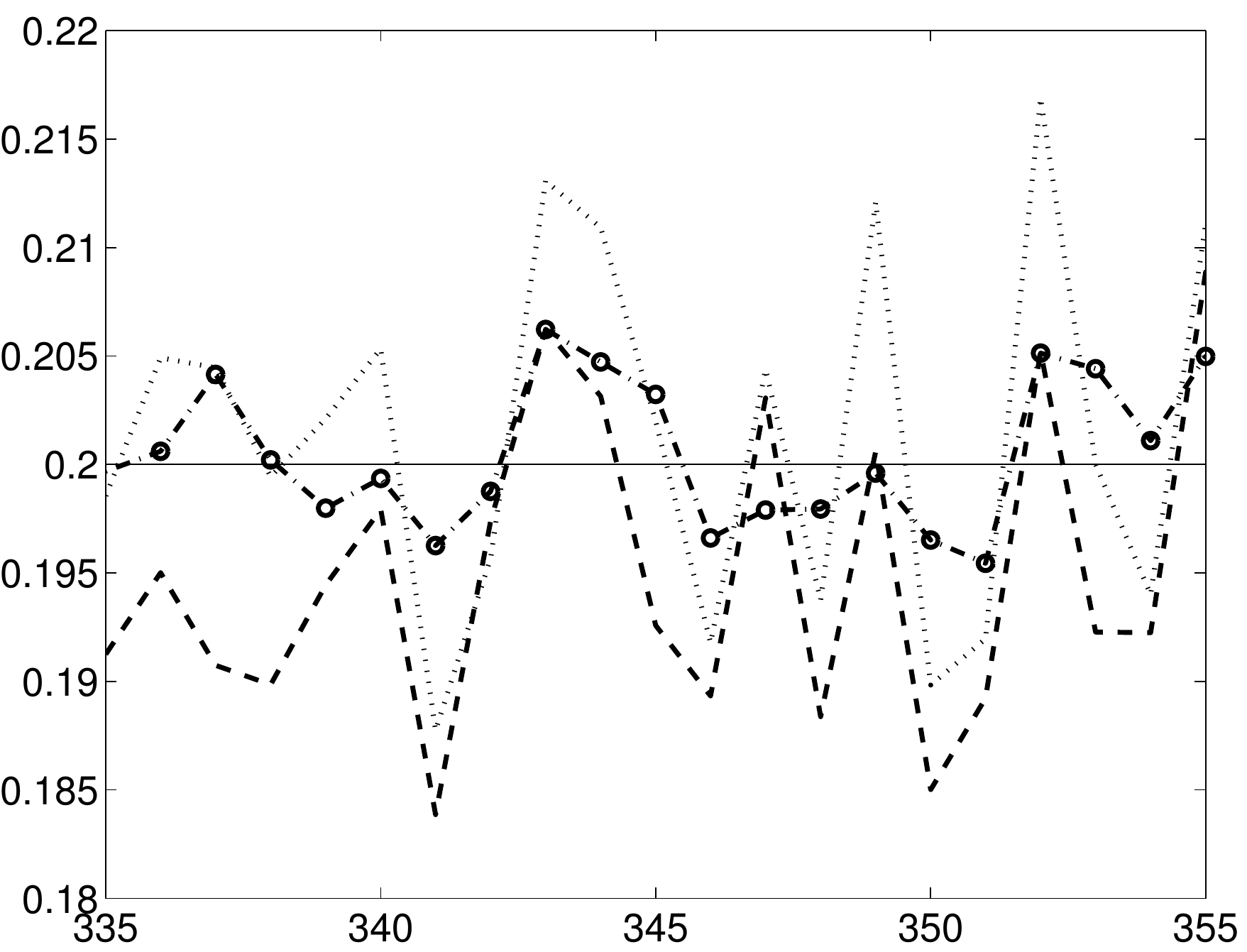}\label{fig:fig10_b}

} 

\caption{The central profiles along the vertical direction of Figure \ref{fig:fig6_a},
\ref{fig:fig6_b}, \ref{fig:fig6_c}. (a) is the zoomed image at the
place close to the edge of different image structures. (b) is the
zoomed image at the place far away from the edge of different image
structures. In the two figures, the solid line corresponds to the
phantom; the dashed line is corresponding to NIRM; the dotted line
corresponds to TIRM; the dash-dot line with circle marks corresponds
to SIRM. }

\label{fig:fig10}
\end{figure}

Table \ref{tab:1} tells that the TIRM has the smallest distance.
However, TIRM has an over correction at the image edges. The over
correction can be seen in Fig.~\ref{fig:fig10_a}. Here the dotted
line is far away from the solid line compared to dashed line and dash-dot
line. The dashed line and dash-dot line are close to each other. Fig.~\ref{fig:fig10_a}
shows that TIRM has a over correction at the image edges. The areas
close to the edge of the different image structures are strongly relayed
to the distance defined in Eq.(\ref{eq:210}). This kind of over correction
can reduce the distance, but it causes the reconstructed image to
be oscillated at the image edges. In a clinical cases the over correction
can not be accepted. The over correction is easy to be thought as
some kind of real structure, it is dangerous to clinical situation.
Even though TIRM has the smallest distance, the reconstructed image
through TIRM is noisier than other two algorithms. TIRM is rarely
used directly in clinics. The reference \cite{JohanSunnegardh} is
the example of indirectly using TIRM. It is TIRM plus pre-filtering
and post-filtering in the reconstruction. Pre-filtering can cause
the lose of information. 

In contrast, the SIRM reduces the oscillation at the place close to
the edges and reduces the artifacts at the place far away from the
edges simultaneously, which can be seen in Figure~\ref{fig:fig10_b}.
Here the dash-dot line is the closest line to the solid line. The
dash-dot line is corresponding to SIRM.

According to the above discussion, SIRM and LIRM have better quality
compared with NIRM and TIRM in image reconstruction with FFOV.

\section{Conclusions and future work}

Two generalized iterative refinement methods LIRM and SIRM have been
introduced. As an example, simple inverse problem to\emph{ }utilize
LIRM has been given. The LIRM eliminates the over correction and it
is less noise sensitive comparing to TIRM.

SIRM has been applied to the CT image reconstruction from untruncated
parallel-beam projections. The simulations shown that it can reduce
the normal artifacts remarkably, which exists in the reconstruction
with FBP algorithm. SIRM has been compared to the TIRM and NIRM. The
result shows that the SIRM has less artifacts in reconstructed image
and TIRM is more sensitive to noises. The distance of SIRM is smaller
than NIRM. The smallest distance is obtained through TIRM. However
the smallest distance is achieved through an over-correction in the
places close to the image edges, which can not be accepted. 

These authors have shown that LIRM is a special case of SIRM. NIRM
and TIRM are special cases of LIRM. Hence LIRM and SIRM are two generalized
iterative refinement reconstruction methods. SIRM and LIRM can be
seen as local inverse applied to the image reconstruction of full
field of veiw. Hence, SIRM and LIRM can be seen as generalized iterative
refinement method(GIRM) and local inverse method for FFOV. SIRM and
LIRM do not minimize the noise or artifacts alone but minimize the
total values of the noise and artifacts. Even the SIRM and LIRM are
developed in the field of CT image reconstruction, these authors believe
they are a general methods and can be applied widely in physics and
applied mathematics where IRM can be applied. 

The future work of these authors is to implement the LIRM in CT image
reconstruction. these authors also plan to implement SIRM and LIRM
in fan-beam and cone-beam geometries.

\end{document}